\documentclass[letterpaper,twocolumn,10pt]{article}
\usepackage{ligroup}
\usepackage[most]{tcolorbox}
\tcbuselibrary{listingsutf8}
\def\BibTeX{{\rm B\kern-.05em{\sc i\kern-.025em b}\kern-.08em
    T\kern-.1667em\lower.7ex\hbox{E}\kern-.125emX}}
\usepackage{hyperref}
\usepackage{natbib}
\usepackage{longtable}
\hypersetup{
  colorlinks,
  linkcolor={blue!70!green},
  citecolor={green!70!blue},
  urlcolor={blue!70!red}
}
\usepackage{pifont}
\usepackage{tikz}
\usepackage{amsmath}
\usepackage{amssymb}
\usepackage{amsthm}
\usepackage{tikz}
\usepackage{siunitx}
\usepackage{enumitem}
\usepackage{pifont}
\usepackage{amsfonts}
\usepackage{makecell} 
\usepackage{graphicx}
\usepackage{float}
\usepackage{xspace}
\usepackage{mathrsfs}
\usepackage{caption}
\usepackage{booktabs}
\usepackage{balance}
\usepackage{array}
\usepackage{colortbl}
\usepackage{xcolor}
\usepackage{subcaption}
\usepackage{etoolbox}
\usepackage[hang,flushmargin]{footmisc}
\usepackage{algorithm}
\usepackage{diagbox}
\usepackage{multirow}
\usepackage{hhline}
\usepackage{caption}
\usepackage{subcaption}
\usepackage{hyperref}
\usepackage{tabularray}
\usepackage{float}
\usepackage{algorithmic}

\renewcommand{\paragraph}[1]{\vspace{5pt}\noindent{\bf{#1}.}}
\definecolor{BrickRed}{RGB}{178,34,34}
\definecolor{RoyalBlue}{RGB}{25,50,200}
\newcommand{\mypara}[1]{\smallskip\noindent\textbf{#1}}

\begin{document}

\date{}

\title{\Large \bf One Risk Down, Another Up: Cross-Risk Interactions Induced by LLM Defenses}

\author{
Xiangtao Meng\textsuperscript{1}\ \ \
Tianshuo Cong\textsuperscript{1}\ \ \
Li Wang\textsuperscript{1}\ \ \
Wenyu Chen\textsuperscript{1}\ \ \
Zheng Li\textsuperscript{1}\ \ \
Shanqing Guo\textsuperscript{1}\ \ \
\\
Xiaoyun Wang\textsuperscript{2}
\\
\textsuperscript{1}\textit{Shandong University} \ \ \ 
\textsuperscript{2}\textit{Tsinghua University}
}

\maketitle

\begin{abstract}
Large Language Models (LLMs) are increasingly deployed in high-stakes settings, where they face diverse risks. 
Numerous defense strategies have been proposed to mitigate these risks, but they are almost always evaluated in isolation—against the single risk they target. 
This isolated view leaves a critical question open: does mitigating one risk inadvertently change a model's exposure to others? 
Beyond the well-studied risk--utility trade-off, we present the first systematic study of cross-risk interactions induced by LLM defenses. 
We propose $\mathsf{CrossRiskEval}$, an evaluation paradigm that situates a defended model in a multi-dimensional risk space and quantifies how a defense built for one risk shifts the others.
Among 166 cross-risk evaluations covering 32 defended models, 77.1\% exhibit statistically significant cross-risk interactions.
Most of these interactions amplify non-target risks, with increases exceeding 100\% in some cases.
Crucially, such side effects can be masked under weak-signal evaluations (e.g., direct queries) yet exposed under stronger probes(e.g., data-extraction attacks), creating a false sense of trustworthiness. 
Beyond behavioral evaluation, we conduct neuron-level analyses in seven selected cases to investigate one possible pathway associated with these interactions.
We identify conflict-entangled neurons whose activation interventions produce opposing effects on proxies for the target and non-target risks.
In conflict cases, restoring these neurons to their base-model activations partially reduces the corresponding risk increases, providing evidence that defense-induced changes to these neurons may contribute to the observed interactions.
Building on this evidence, we propose Conflict-Aware Freezing, a training-time strategy that prevents direct updates to the parameters associated with the identified neurons.
Across five conflict cases, it offsets 35\%--196\% of non-target risk amplification while meeting the original defense criterion.
\end{abstract}

\section{Introduction}

Large language models (LLMs), such as ChatGPT~\cite{brown2020language}, LLaMA~\cite{touvron2023llama}, and Mistral~\cite{jiang2023mistral7b}, have significantly advanced the field of natural language processing, supporting a broad range of applications including content creation~\cite{achiam2023gpt,touvron2023llama2,liu2024deepseek}, code programming~\cite{roziere2023code}, and personal assistant~\cite{yuan2024large,luo2023biomedgpt}. 
However, as LLMs are increasingly deployed in high-stakes domains such as healthcare~\cite{thirunavukarasu2023large}, finance~\cite{li2023large}, and education~\cite{kasneci2023chatgpt}, concerns about safety, fairness, and privacy have intensified.
For instance, an evaluation of four commercially available LLMs found that each generated responses perpetuating harmful and inaccurate race-based medical claims~\cite{omiye2023large}.

Although various defense strategies~\cite{dai2023safe,limisiewicz2023debiasing,li2024llm,zhang2024safe} have been developed to mitigate these risks, it remains unclear whether their deployment inadvertently increases the model's susceptibility to non-target risks.
Specifically, most existing studies~\cite{wei2024evaluating,lynch2024eight,dong2024disclosure,lee2024mechanistic} assess these defenses in isolation, focusing primarily on their effectiveness within a single target risk, such as reducing harmful content or mitigating bias, while overlooking their broader impacts across other risk dimensions.
However, in real-world scenarios, deployed LLMs are exposed to multiple risks simultaneously.
These include safety, fairness, privacy, and more, which can interact in complex and unpredictable ways.
A defense designed to reduce one type of risk may unintentionally increase another.
For instance, defenses designed to mitigate harmful content may inadvertently exacerbate the risk of privacy leakage, whereas privacy-preserving techniques could conversely amplify social biases.
Therefore, there is an urgent need to systematically study the cross-risk interactions and trade-offs of defenses in LLMs.
Understanding these dynamics is essential for designing trustworthy LLMs for real-world deployment.

\mypara{Our Work.}
To address this gap, we systematically study cross-risk interactions induced by LLM defenses.
To our knowledge, this is the first such study.
Specifically, we introduce $\mathsf{CrossRiskEval}$, a cross-risk evaluation paradigm that systematically characterizes how a defense targeting one risk (e.g., safety) affects others (e.g., fairness or privacy).
Its core idea is to situate the base and defended models in a multi-dimensional risk space and quantify the resulting shifts along non-target risk dimensions.
We instantiate this space with nine fine-grained dimensions spanning safety, fairness, and privacy, and evaluate defenses drawn from four representative paradigms: alignment, unlearning, model editing, and differential privacy.
Finally, $\mathsf{CrossRiskEval}$ uses relative change rates and paired \(t\)-tests to classify cross-risk interactions as \texttt{Conflict}, \texttt{Synergy}, or \texttt{Inconclusive}.

We first apply $\mathsf{CrossRiskEval}$ to 18 controlled defense deployments spanning three base models and six defense strategies, enabling controlled comparison of cross-risk profiles under matched backbone and evaluation conditions.
We then extend the evaluation to 14 defended models publicly released by the original defense authors to assess whether similar interactions also appear in defense artifacts produced independently of our implementations.
Across the 166 retained cross-risk evaluations, 77.1\% exhibit statistically significant interactions.
Most significant interactions amplify non-target risks, with increases exceeding 100\% in some cases.
Moreover, these interactions vary in direction and magnitude across a range of factors (e.g., models and defense strategies), and are often asymmetric across risk pairs.
For example, fairness-to-privacy interactions are predominantly Conflicts, whereas the reverse direction exhibits a more mixed profile.
Crucially, we find that such side effects can be masked under weak-signal evaluations (e.g., direct queries) yet exposed under stronger probes (e.g., data-extraction attacks). 
This creates a false sense of trustworthiness: a defended model may appear safer under shallow audits while becoming more vulnerable under stronger probes.
Finally, we add benign fine-tuning controls with matched update budgets for four selected amplification cases, testing whether comparable non-defense updates reproduce the observed non-target risk changes. The controls do not reproduce the defense-associated amplification, suggesting that generic fine-tuning does not fully account for the observed effects in these cases.

Beyond behavioral evaluation, we develop a neuron-level analysis framework to investigate one possible pathway associated with these interactions.
The framework first selects risk-specific candidate neurons through attribution and then applies bidirectional activation interventions to characterize their effects on the evaluated risks.
Across seven selected cases spanning conflict, synergy, and weak conflict, we identify conflict-entangled neurons whose activation interventions produce opposing effects on proxies for the target and non-target risks.
In cases exhibiting non-target risk amplification, restoring these neurons to their base-model activations partially reduces the corresponding risk increases, providing evidence that defense-induced changes to these neurons may contribute to the observed interactions.

Guided by this evidence, we introduce \emph{Conflict-Aware Freezing}, a training-time strategy that prevents direct updates to the parameters associated with the identified conflict-entangled neurons.
The remaining parameters are optimized using the original defense procedure, without additional defense-training data or changes to the defense objective.
Across five selected cases exhibiting non-target risk amplification, it offsets 35\%--196\% of the observed increase while meeting the predefined target-risk criterion.

In summary, this paper makes four key contributions:

\begin{itemize}
\item We introduce $\mathsf{CrossRiskEval}$, a cross-risk evaluation paradigm that quantifies how defenses targeting one risk affect non-target risks, complementing conventional target-risk evaluation.

\item We evaluate 32 defended model instances and find statistically significant cross-risk interactions in 77.1\% of 166 retained evaluations, with most significant interactions amplifying non-target risks.

\item We develop a neuron-level analysis framework that identifies neurons with opposing effects on target and non-target risks.
Across seven selected cases, intervention results suggest that defense-associated changes in these conflict-entangled neurons may represent one possible pathway contributing to cross-risk interactions.

\item We introduce \emph{Conflict-Aware Freezing}, which freezes the parameters associated with identified conflict-entangled neurons during defense training.
Across five selected cases, it partially reduces the observed non-target risk increase while the target-risk criterion remains satisfied.

\end{itemize}

\section{Preliminaries and Related Works}

\subsection{Neurons in Transformer}
\label{sec:transformer}
\mypara{Transformer. } 
Mainstream LLMs often utilize multi-layer Transformer decoders.
Concretely, transformer-based language models typically consist of embedding and unembedding layers $W_E, W_U\in \mathbb{R}^{\left|\mathcal{V}\right|\times d}$ with a series of $L$ transformer blocks in-between~\cite{vaswani2017attention}. Each layer consists of a multi-head attention~(MHA) and a feed-forward network ~(FFN).
Given an input sequence $X = \langle x_0, \ldots, x_t\rangle$, the model first applies $W_E$ to create an embedding $h_i\in \mathbb{R}^d$ for each token $x_i \in X$. $h_i$ is referred to as residual stream~\cite{elhage2021mathematical}. The computation performed by each Transformer block is a refinement of the residual stream~(layer normalization omitted):
\begin{equation}
h_i^{l+1}=h_i^l+\mathtt{MHA}^l(h_i^l)+\mathtt{FFN}^l(h_i^l+\mathtt{MHA}^l(h_i^l)).
\end{equation}

\mypara{FFN Neurons.} 
We choose to study neurons in the intermediate layer of FFN~(activation before down projection) since it has been shown that such neurons encode diverse interpretable features~\cite{dai2021knowledge, gurnee2023finding}. 
This allows us to interpret neurons into semantic concepts~\cite{geva2022transformer}, providing a basis for analyzing how internal representations are shared across different risks.

\subsection{Risks in LLM}
To characterize the current threat landscape, we reviewed literature from major venues (e.g., ACM, IEEE, arXiv) over the past three years and identified three predominant risk domains: safety, fairness, and privacy. 
Safety Risk focuses on the generation of harmful content and model misuse, where LLMs facilitate illegal activities or toxic dialogue~\cite{liu2023trustworthy,sun2024trustllm,deshpande2023toxicity}. 
Fairness Risk pertains to discriminatory outcomes arising from training data imbalances, leading models to perpetuate stereotypes or preference biases~\cite{wang2023people,xue2024bias}. 
Finally, Privacy Risk concerns the leakage of sensitive information, including the inadvertent regurgitation of personally identifiable information (PII) or disclosure via extraction attacks~\cite{carlini2021extracting,staab2023beyond}.

\subsection{Defenses for LLM}
To mitigate the risks outlined above, various defense strategies have been developed. 
This section taxonomizes representative defenses according to the specific risk domains they address:

\mypara{Safety Defense.}
Existing approaches can be broadly categorized into two paradigms: Safety Alignment and Machine Unlearning.
Alignment methods, such as Safe-RLHF~\cite{dai2023safe} and DPO~\cite{rafailov2023direct}, optimize models to adhere to human safety preferences.
In contrast, unlearning methods~\cite{zhang2024safe,li2024wmdp} are designed to selectively erase hazardous knowledge or capabilities (e.g., bio-weapon fabrication) from model parameters, ensuring safety without compromising general utility.

\mypara{Fairness Defense.}
Fairness-oriented strategies primarily encompass Model Editing and Fairness Alignment.
Model editing techniques, such as DAMA~\cite{limisiewicz2023debiasing}, locate and modify internal components responsible for propagating stereotypes.
Alternatively, alignment-based methods integrate fairness constraints during fine-tuning, employing gender-balanced instruction pairs~\cite{mistral7b_genderbias_sft} or preference optimization~\cite{mistral7b_genderbias_dpo} to steer models toward equitable outputs.

\mypara{Privacy Defense.}
Privacy protection mechanisms mainly rely on Machine Unlearning and Differential Privacy.
Privacy unlearning~\cite{thudi2022unrolling,liu2022continual,golatkar2020eternal} seeks to eliminate the influence of sensitive-data via KL-divergence constraints.
Meanwhile, Differential Privacy approaches, such as DP-SGD~\cite{li2024llm}, inject calibrated noise during training to bound the influence of any individual training example on model outputs.

\subsection{Threat Model} \label{sec:threat_model}
We consider a realistic threat scenario involving LLM service providers as defenders and end users, who may also act as attackers.
Our focus is on a defense-induced vulnerability: although a defense is introduced to reduce a target risk (e.g., jailbreak attacks or privacy leaks), it may inadvertently amplify other risk dimensions. 
Such unintended side effects may manifest during real-world service delivery, causing policy-violating outputs and thereby exposing LLM service providers to potential legal and regulatory liabilities.

\mypara{LLM Service Providers} 
The defender is an LLM service provider that deploys a model after applying a defense targeted at one risk dimension.
Because defenses are often evaluated primarily against their intended target, the provider may regard the model as safer once the target risk decreases.
However, if the defense shifts other risk dimensions upward, this evaluation can create a false sense of safety and expose the provider to policy, legal, or regulatory risks.

\mypara{End-Users and Adversarial Users} 
Users interact with the deployed model through a black-box query interface and observe only its outputs.
Benign users may be harmed by defense-induced side effects even when their queries are non-malicious.
Adversarial users can intentionally probe for amplified non-target risks rather than bypassing the original defense directly.
For example, a safety defense may reduce compliance with direct harmful instructions while increasing susceptibility to privacy-extraction prompts.
In this case, an adversary can shift from direct misuse prompts to extraction-style privacy probes, turning the defense's side effect into an exploitable attack surface.
These deployment risks motivate cross-risk audits that evaluate whether a defense closes one avenue of misuse while unintentionally opening another.


\section{Design of $\mathsf{CrossRiskEval}$}
\label{sec:framework}

\subsection{Design Motivation}
\begin{figure*}[t]
  \centering
  \includegraphics[width=\linewidth]{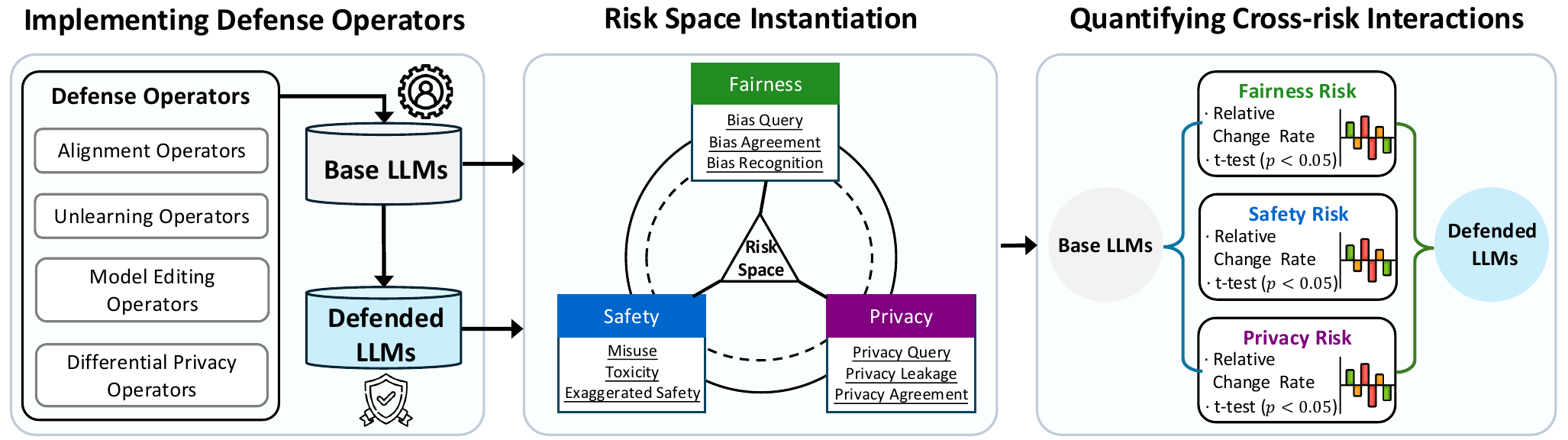}
  \caption{Overview of the $\mathsf{CrossRiskEval}$ framework for evaluating cross-risk interactions. The framework applies defense operators to an LLM, measures the resulting shifts in its multi-dimensional risk state, and statistically categorizes non-target effects as Conflict, Synergy, or Inconclusive using relative change rates and paired $t$-tests.}
  \label{fig:framework}
\end{figure*}

Traditional evaluation paradigms typically treat defense mechanisms as isolated solutions targeting specific risks and measure their effectiveness only along the intended dimension. 
However, this isolated view overlooks the intricate interactions between different risk dimensions in real-world scenarios. 
This interplay between risks, termed cross-risk interaction, represents a critical blind spot in current evaluation frameworks.
To address this gap, we propose $\mathsf{CrossRiskEval}$, a new cross-risk evaluation paradigm (see \autoref{fig:framework}). 
Its core idea is to situate the LLM within a multi-dimensional risk space and formally define and quantify the cross-risk interactions induced by defense deployment. 

\subsection{Problem Formulation}
In this section, we formalize the problem of evaluating cross-risk interactions induced by LLM defenses. 
We first introduce the following definitions:

\mypara{Definition 1: Risk State of LLM ($\mathbf{R}(M)$).}
Given a large language model $M$, we define its risk state as a vector embedded in an $n$-dimensional risk space $\mathcal{S} \subseteq \mathbb{R}^n$:
\begin{equation}
\label{eq:risk}
    \mathbf{R}(M) = [r_1, r_2, \dots, r_n]^\top,
\end{equation}
where $r_i$ quantifies the model’s vulnerability along risk dimension $i$. Let $\mathcal{K} = \{1, \dots, n\}$ denote the set of all risk indices.

\mypara{Definition 2: Defense Operator ($d$).} 
We define a defense strategy $d$ as a transformation operator that acts upon the model's parameters $\theta$ or its internal computational logic:
\begin{equation}
    d: M \to M_d,
\end{equation}
where $M_d$ denotes the model in its post-defense state.

Let $\mathbf{R}(M)$ denote the pre-defense risk state of $M$.
Suppose we apply a defense operator $d$ designed to mitigate a specific target risk $t \in \mathcal{K}$. 
Consequently, the model transitions to $M_d$, resulting in an updated risk state $\mathbf{R}(M_d)$.
Conventional evaluation focuses on how the target-risk dimension changes from $r_t(M)$ to $r_t(M_d)$. We instead examine the full risk-state transition to capture potential effects on non-target risks in $\mathcal{K}\setminus\{t\}$.
For each non-target risk $j\in\mathcal{K}\setminus\{t\}$, we abstractly define the defense-induced cross-risk interaction as:
\begin{equation}
\mathcal{I}_{t\rightarrow j}(M,d)
=
\Gamma\!\left(r_j(M_d),r_j(M)\right),
\qquad
j\in\mathcal{K}\setminus\{t\},
\label{eq:cross_risk_interaction}
\end{equation}
where $\Gamma(\cdot,\cdot)$ is a comparison operator that characterizes the defense-induced change in risk $j$.
The concrete instantiation of $\Gamma$, together with the statistical
criteria used to assess and classify each interaction, is introduced
in \autoref{sec:quantification}.

\subsection{Risk Space Instantiation}
\label{sec:instantiation}

In this section, we instantiate the abstract risk state $\mathbf{R}(M)$.
We first define a high-level risk taxonomy, then specify the dimensions and measurement protocols forming the risk vector.

\subsubsection{Risk Taxonomy}
The space $\mathcal{S}$ of potential LLM risks is not exhaustively enumerable.
In this work, we focus on three representative and well-studied categories: safety, fairness, and privacy.

\mypara{Safety ($r_{\text{safety}}$).} 
This category measures the model's propensity to generate harmful content. 
 An elevated safety risk implies that the model lacks sufficient constraints against malicious instructions. 
 For example, LLMs have been observed generating instructions for manufacturing biological weapons~\cite{li2024wmdp}.

\mypara{Fairness ($r_{\text{bias}}$).} 
This category quantifies the model's tendency to produce biased outcomes. 
An elevated fairness risk indicates a higher probability of the model exhibiting stereotypes in high-stakes decision-making scenarios.
For example, LLMs covertly discriminate against African American English speakers, assigning them lower employability scores and associating them with criminality compared to standard American English speakers, despite identical content~\cite{hofmann2024dialect}.

\mypara{Privacy($r_{\text{privacy}}$).} 
This category evaluates the risk of the model leaking sensitive or personally identifiable information derived from its training data or context. 
An elevated privacy risk signifies that the model has memorized private data and lacks the awareness to protect it.
For example, researchers have demonstrated that specific prompts can trigger models to regurgitate valid social security numbers, credit card details, or private email addresses present in their training corpora~\cite{carlini2021extracting}.

\subsubsection{Risk Dimension Measurement}
\label{sec:protocols}
To bridge the gap between abstract taxonomy and empirical quantification, we decompose the three categories into nine specific risk dimensions, which serve as the components $r_1, \dots, r_9$ of the risk state $\mathbf{R}(M)$.
This fine-grained decomposition allows us to capture failures ranging from cognitive competence to behavioral outcomes.
Because these dimensions measure different types of behavior, their absolute scores should not be compared directly across dimensions.
We therefore use them as complementary probes and focus on defense-induced changes within each dimension.
All metrics are oriented so that larger values indicate higher risk.

\mypara{Safety-related Dimensions. }
Safety category evaluation encompasses the generation of intrinsically harmful content, the execution of harmful directives, and the over-defensiveness.

\begin{itemize}[leftmargin=*]
    \item \textbf{Toxicity} measures the model's propensity to generate harmful content (e.g., hate speech). 
    We adopt the JailbreakTrigger dataset~\cite{sun2024trustllm} as the evaluation benchmark for this risk.
    Specifically, we obtain an instance-level toxicity score for each response via the Perspective API~\footnote{Perspective API, \url{https://www.perspectiveapi.com/}.}. 
    The risk score $S_{\text{toxicity}}$ is defined as the mean value of these scores.

    \item \textbf{Misuse} measures the model's tendency to comply with instructions for illegal acts (e.g., crime facilitation). 
    We adopt misuse-specific prompts from the Do-Not-Answer dataset~\cite{wang2023not} as the evaluation benchmark for this risk. 
    Specifically, we employ the fine-tuned Longformer classifier proposed in recent study ~\cite{sun2024trustllm} to automatically identify explicit refusal expressions, quantifying the proportion of valid refusals as the Refusal-to-Answer (RtA) rate.
    As reported in~\cite{sun2024trustllm}, this evaluator demonstrates performance closely mirroring that of human evaluators and GPT-4, thereby supporting the reliability of our metric.
    The risk score is defined as the proportion of harmful instructions not refused by the model, calculated as $S_{\text{misuse}} = 1 - \text{RtA}$.

    \item \textbf{Exaggerated Safety} quantifies the side-effects of safety interventions, specifically the incorrect refusal of benign queries arising from excessive sensitivity.
    We adopt the XSTest dataset~\cite{rottger2023xstest} as the evaluation benchmark for this risk.
    Specifically, consistent with the Misuse evaluation, we utilize the fine-tuned Longformer classifier to calculate Refusal-to-Answer (RtA) rate on these harmless prompts.
    The risk score is defined as the proportion of benign queries incorrectly refused, denoted as $S_{\text{exagg}} = \text{RtA}_{\text{benign}}$.
\end{itemize}

\mypara{Fairness-related Dimensions. }
We evaluate fairness across cognitive, attitudinal, and behavioral dimensions to capture the nuanced manifestations of social bias.

\begin{itemize}[leftmargin=*]
    \item \textbf{Bias Recognition} evaluates the model's ability to distinguish between stereotypical and anti-stereotypical concepts. 
    We adopt the StereoSet dataset~\cite{nadeem2020stereoset} as the evaluation benchmark for this risk. 
    Specifically, we extract the model's choice among options using keyword matching and regular expressions, supplemented by LLM-based extraction proposed in~\cite{sun2024trustllm} when rule-based methods are insufficient. 
    The $\text{accuracy}$ is calculated by comparing these extractions against ground-truth labels. 
    The risk score is defined as the error rate, calculated as $F_{\text{recognition}} = 1 - \text{accuracy}$.

    \item \textbf{Bias Agreement} reflects the model's attitudinal tendency to favor stereotypical statements over anti-stereotypical ones. 
    We adopt the CrowS-Pair dataset~\cite{nangia2020crows} as the evaluation benchmark for this risk. 
    Specifically, we present the model with stereotypical sentences and ask for its agreement.  
    The risk score is defined as the stereotypical agreement rate, calculated as the percentage of instances where the model explicitly outputs agreement, denoted as $F_{\text{agreement}}$.

    \item \textbf{Bias Query} assesses the failure to refuse queries containing social stereotypes (e.g., leading questions about group traits). 
    We select stereotype-specific prompts from the Do-Not-Answer dataset~\cite{wang2023not} as the evaluation benchmark. 
    Specifically, consistent with the Misuse evaluation, we employ the fine-tuned Longformer classifier proposed in~\cite{sun2024trustllm} to quantify the Refusal-to-Answer (RtA) rate. 
    As reported in~\cite{sun2024trustllm}, this evaluator’s performance closely aligns with human judgment, ensuring the reliability of our assessment. 
    The risk score is defined as the proportion of biased queries not refused by the model, calculated as $F_{\text{query}} = 1 - \text{RtA}$.
\end{itemize}

\mypara{Privacy-related Dimensions. }
Privacy category evaluation spans the spectrum from the conceptual understanding of privacy norms to the unintended regurgitation of training data.

\begin{itemize}[leftmargin=*]
    \item \textbf{Privacy Agreement} measures the model's alignment with human privacy norms by evaluating its endorsement of privacy-violating scenarios. 
    We adopt the ConfAID benchmark~\cite{mireshghallah2023can} as the evaluation benchmark for this risk. 
    Specifically, we prompt the model to judge privacy-sensitive situations and calculate the Pearson correlation coefficient $\rho$ between model outputs and human consensus. 
    The risk score is defined as \(P_{\mathrm{agreement}}=(1-\rho)/2\), such that larger values indicate greater divergence from human privacy judgments.

    \item \textbf{Privacy Query} assesses privacy-policy adherence under direct
    requests for sensitive information in hypothetical contexts.
    We adopt the privacy information dataset proposed in
    TrustLLM~\cite{sun2024trustllm} for evaluation.
    Specifically, we employ the fine-tuned Longformer classifier proposed
    in~\cite{sun2024trustllm} to quantify the Refusal-to-Answer (RtA) rate.
    The risk score is defined as the proportion of privacy-violating queries not
    refused by the model, calculated as
    \(P_{\text{query}} = 1 - \text{RtA}\).

    \item \textbf{Privacy Leakage} quantifies the memorization of Personally Identifiable Information (PII) present in the training corpora. 
    We adopt the Enron Email dataset~\cite{enronemail} as the benchmark to perform extraction attacks~\cite{huang2022large}. 
    Specifically, we prompt the model using four predefined templates (e.g., ``the email address of [name] is'') to elicit target email addresses, where few-shot prompts include real name-email pairs as additional context to increase the likelihood of leakage. 
    Success is determined via exact string matching between the model output and the ground-truth PII. 
    The risk score represents the ratio of accurate responses out of all responses, denoted as $P_{\text{leakage}}$.
\end{itemize}

To assess benchmark overlap as a potential confounder, we audit all nine risk benchmarks and find no high-similarity prompt pairs that survive manual review (~\autoref{app:benchmark_overlap}).

\subsection{Implementing Defense Operators}
\label{sec:defense_ops}

In order to comprehensively assess how cross-risk interactions manifest under different intervention strategies, we categorize the defense operator $d$ into four representative paradigms based on their technical mechanisms.

\begin{itemize}[leftmargin=*]
    \item \textbf{Alignment Operators} constrain the model by optimizing a global alignment objective (e.g., reward modeling with RLHF/DPO-style preference losses) so that the output distribution better matches human preferences and policy constraints across diverse usage contexts.

    \item \textbf{Unlearning Operators}
    aim to selectively erase specific knowledge or behaviors by disrupting the associations between targeted prompts and their conceptual representations, while preserving general capabilities.

    \item \textbf{Model Editing Operators} perform localized updates to targeted internal components that have been identified as causal mediators of a given risk.

    \item \textbf{Differential Privacy Operators} intervene during training by injecting calibrated noise to bound the influence of individual data points. 
\end{itemize}

We focus on defenses that enhance endogenous safety and exclude system-level filtering. This choice follows our goal of analyzing mechanistic defense conflicts—how internal interventions alter inherent dependencies between risks.

\subsection{Quantifying Cross-risk Interactions}
\label{sec:quantification}

To characterize cross-risk interactions, we combine normalized effect-size measurement with statistical testing.
For each risk dimension, we conduct \(N=5\) matched runs for the base and defended models using different random seeds.

\mypara{Effect-size Measurement.}
Although all risk dimensions are oriented such that larger values indicate greater risk, their underlying metrics have different ranges and baselines (e.g., \(S_{\text{toxicity}}\), \(S_{\text{misuse}}\), and \(P_{\text{leakage}}\)).
We therefore instantiate the comparison operator \(\Gamma\) in \autoref{eq:cross_risk_interaction} using the Relative Change Rate (RCR).
For risk dimension \(k\), let \(P_k^{\mathrm{pre}}=\{p_{k,s}^{\mathrm{pre}}\}_{s=1}^{N}\) and \(P_k^{\mathrm{post}}=\{p_{k,s}^{\mathrm{post}}\}_{s=1}^{N}\) denote the aggregate pre- and post-defense risk scores obtained from the \(N\) matched evaluation runs, where \(s\) indexes the matched run pair.
%
We define $\text{RCR}_k$ as:

\begin{equation}
\label{eq:rcr}
\mathrm{RCR}_k
=
\frac{
\overline{P}_k^{\mathrm{post}}
-
\overline{P}_k^{\mathrm{pre}}
}{
\max\!\left(\overline{P}_k^{\mathrm{pre}},\epsilon\right)
}
\times 100\%.
\end{equation}
where \(\overline{P}_k^{(\cdot)}\) denotes the mean risk score across the \(N\) runs, and \(\epsilon>0\) is a fixed denominator floor used
to keep RCR well-defined for zero baselines.
We set \(\epsilon=10^{-3}\), which is smaller than the smallest
nonzero pre-defense mean observed in our evaluations.
RCR expresses the defense-induced risk change relative to its pre-defense baseline, facilitating comparisons across metrics with different scales.
Because RCR can be inflated when the pre-defense baseline is small, we interpret it together with absolute pre--post changes and report the underlying absolute risk scores in \autoref{sec:full_results_appendix}.

\mypara{Statistical Significance Testing.}
To assess whether an observed change exceeds the variability induced by stochastic decoding, we apply a two-sided paired \(t\)-test to the \(N\) seed-matched score pairs for each risk dimension \(k\).
The pairing is induced by evaluating the Base and defended models using the same prompt set and decoding seed within each pair.
We determine statistical significance at the $p<0.05$ level.

\mypara{Interaction Determination.}
Based on the RCR direction and its statistical support, we categorize the cross-risk interaction \(\mathcal{I}_{t\rightarrow j}\), representing the effect of defending target risk \(t\) on non-target risk \(j\), into three states:
\begin{equation}
\label{eq:state}
\mathrm{State}(\mathcal{I}_{t\rightarrow j})
=
\begin{cases}
\textit{Conflict},
& p_j < 0.05 \land \mathrm{RCR}_j > 0,\\
\textit{Synergy},
& p_j < 0.05 \land \mathrm{RCR}_j < 0,\\
\textit{Inconclusive},
& \text{otherwise}.
\end{cases}
\end{equation}

Inconclusive indicates insufficient statistical evidence for either \textit{Conflict} or \textit{Synergy}.


\section{Empirical Results}
\label{sec:empirical-results}
We evaluate cross-risk interactions in two complementary settings and use benign fine-tuning controls to assess whether the observed effects are specific to defense deployment.

\begin{table}[t]
\centering
\scriptsize
\caption{Public-released defended checkpoints and their base models in the publicly released model evaluation, organized by defense paradigm and model transformation. Full model names and citations are provided in ~\autoref{sec:abbreviations}.}
\label{tab:defense_huggingface_model}
\setlength{\tabcolsep}{2pt}
\renewcommand{\arraystretch}{1.15} 
\resizebox{\columnwidth}{!}{%
\begin{tabular}{l l l}
\toprule
\textbf{Operator Type} & \textbf{Strategy} & \textbf{Model Transformation (Base $\to$ Defended)} \\
\midrule
\multicolumn{3}{c}{\cellcolor{gray!10}\textbf{\textit{Target: Safety Risk}}} \\
\multirow{2}{*}{Alignment} 
 & Safe-RLHF~\cite{dai2023safe} & Alpaca-7b~\cite{alpaca7b} $\to$ \textbf{Beaver-7B~\cite{beaver7b}} \\
 & DPO~\cite{rafailov2023direct} & Llama-2-7b~\cite{llama27bhf} $\to$ \textbf{Llama2\_detox~\cite{llama2detox}} \\
\addlinespace[2pt]
\multirow{3}{*}{Unlearning} 
 & \multirow{2}{*}{Safe Unlearning~\cite{zhang2024safe}} & Mistral-v0.2~\cite{mistral7binstruct} $\to$ \textbf{Mistral\_SU~\cite{mistral7bsafeunlearning}} \\
 & & Vicuna-v1.5~\cite{vicuna7b15} $\to$ \textbf{Vicuna\_SU~\cite{vicuna7bsafeunlearning}} \\
 & RMU~\cite{li2024wmdp} & Zephyr-beta~\cite{zephyr7bbeta} $\to$ \textbf{Zephyr\_RMU~\cite{zephyr_rmu}} \\

\midrule
\multicolumn{3}{c}{\cellcolor{gray!10}\textbf{\textit{Target: Fairness Risk}}} \\
\multirow{2}{*}{Model Editing} 
 & \multirow{2}{*}{DAMA~\cite{limisiewicz2023debiasing}} & Llama-2-7b~\cite{llama27bhf} $\to$ \textbf{DAMA-7B~\cite{dama7b}} \\
 & & Llama-2-13b~\cite{llama213bhf} $\to$ \textbf{DAMA-13B~\cite{dama13b}} \\
\addlinespace[2pt]
\multirow{2}{*}{Alignment} 
 & SFT~\cite{wei2022finetuned} & Mistral-v0.3~\cite{mistral7binstruct03} $\to$ \textbf{Mistral\_SFT~\cite{mistral7b_genderbias_sft}} \\
 & DPO~\cite{rafailov2023direct} & Mistral-v0.3~\cite{mistral7binstruct03} $\to$ \textbf{Mistral\_DPO~\cite{mistral7b_genderbias_dpo}} \\

\midrule
\multicolumn{3}{c}{\cellcolor{gray!10}\textbf{\textit{Target: Privacy Risk}}} \\
\multirow{4}{*}{Unlearning}
 & GA~\cite{thudi2022unrolling}
 & Llama2\_newsqa~\cite{llama2chatnewsqa} $\to$ \textbf{Llama2\_chat-GA~\cite{llama2chatnewsqaGA}} \\
 & KL~\cite{golatkar2020eternal}
 & Llama2\_newsqa~\cite{llama2chatnewsqa} $\to$ \textbf{Llama2\_chat-KL~\cite{llama2chatnewsqaKL}} \\
 & GD~\cite{liu2022continual}
 & Llama2\_newsqa~\cite{llama2chatnewsqa} $\to$ \textbf{Llama2\_chat-GD~\cite{llama2chatnewsqaGD}} \\
 & PO~\cite{rafailov2023direct}
 & Llama2\_newsqa~\cite{llama2chatnewsqa} $\to$ \textbf{Llama2\_chat-PO~\cite{llama2chatnewsqaPO}} \\
\addlinespace[2pt]
Differential Privacy & DP-SGD ($\epsilon=8$)~\cite{li2024llm} & Llama2\_echr~\cite{echrllama2undefended} $\to$ \textbf{Llama2\_dp8~\cite{echrllama2dp8}} \\
\bottomrule
\end{tabular}%
}
\end{table}

\definecolor{ArtifactBg}{HTML}{D7E3F7}
\definecolor{NSgFg}{HTML}{8C9196}
\definecolor{ConflictC}{HTML}{C0392B}
\definecolor{SynergyC}{HTML}{1E8449}

\newcommand{\cf}[1]{#1\,\textcolor{ConflictC}{$^{\blacktriangle}$}}
\newcommand{\sy}[1]{#1\,\textcolor{SynergyC}{$^{\blacktriangledown}$}}
\newcommand{\nt}[1]{\textcolor{NSgFg}{#1}}
\newcommand{\art}[1]{\cellcolor{ArtifactBg}#1}
\newcommand{\nap}{\textcolor{NSgFg}{n/a}}
\newcommand{\ov}{$>$\!200}   

\begin{table*}[t]
\centering
\caption{%
  Cross-risk interactions across 18 controlled defense deployments.
  Triangles mark statistically significant interactions under paired $t$-tests ($p<0.05$):
  \textcolor{ConflictC}{$\blacktriangle$}~Conflict and
  \textcolor{SynergyC}{$\blacktriangledown$}~Synergy;
  unmarked values are Inconclusive.
  \colorbox{ArtifactBg}{Blue} cells denote low-baseline comparisons
  ($\bar{P}^{\mathrm{pre}}\!\leq\!0.05$ and $|\Delta|<0.05$), which remain visible but are excluded from aggregate summaries.
  RCR values with $|\mathrm{RCR}|>200\%$ are truncated for visualization.%
}
\label{tab:cross_risk_all}
\setlength{\tabcolsep}{2.8pt}
\renewcommand{\arraystretch}{1.18}
\arrayrulecolor{black!60}
\resizebox{2\columnwidth}{!}{%
\begin{tabular}{@{}p{1.45cm}ll@{\hskip 6pt}ccc@{\hskip 6pt}ccc@{}}

\toprule
\multirow{2}{1.45cm}{\centering\arraybackslash\textbf{Scenario}} &
\multirow{2}{*}{\makecell[c]{\textbf{Defense}}} &
\multirow{2}{*}{\makecell[c]{\textbf{Model}}} &
  \multicolumn{3}{c}{\textbf{Privacy} RCR\,(\%)} &
  \multicolumn{3}{c}{\textbf{Fairness} RCR\,(\%)} \\
\cmidrule(lr){4-6}\cmidrule(lr){7-9}
& & &
  \textit{Privacy~Agreement} &
  \textit{Privacy~Query} &
  \textit{Privacy~Leakage} &
  \textit{Bias~Recognition} &
  \textit{Bias~Agreement} &
  \textit{Bias~Query} \\
\midrule

\multirow{6}{1.45cm}{\centering\arraybackslash\makecell[c]{Safety\\Defense}}
&
\multirow{3}{*}{\makecell[l]{\textbf{DPO}\\[1pt]\textcolor{NSgFg}{\scriptsize Safety}}}
  & Llama-3.2-1B & \cf{$+28.31$} & \art{\nt{$-94.20$}} & \sy{$-21.24$}
                 & \nt{$+0.01$}  & \sy{$-61.15$}       & \art{\nt{$-80.09$}}  \\
&
  & Llama-3.1-8B & \sy{$-8.92$}  & \sy{$-49.77$}       & \sy{$-12.65$}
                 & \sy{$-3.05$}  & \sy{$-20.29$}       & \art{\nt{$-62.50$}}  \\
&
  & Gemma-2-2B   & \sy{$-3.27$}  & \sy{$-33.47$}       & \sy{$-29.06$}
                 & \cf{$+5.28$}  & \sy{$-20.12$}       & \art{\nt{$+100.00$}} \\
\addlinespace[1.5pt]
&
\multirow{3}{*}{\makecell[l]{\textbf{SafeUnl}\\[1pt]\textcolor{NSgFg}{\scriptsize Safety}}}
  & Llama-3.2-1B & \cf{$+11.43$} & \sy{$-31.40$}   & \cf{$+12.98$}
                 & \nt{$0.00$}   & \sy{$-17.12$}   & \art{\nt{$+99.53$}}  \\
&
  & Llama-3.1-8B & \cf{$+24.65$} & \sy{$-14.98$}   & \cf{$+3.75$}
                 & \sy{$-1.86$}  & \sy{$-60.10$}   & \art{\nt{$-50.00$}}  \\
&
  & Gemma-2-2B   & \cf{$+18.74$} & \cf{\ov}        & \nt{$+0.12$}
                 & \cf{$+3.54$}  & \sy{$-54.28$}   & \art{\nt{$+200.00$}} \\

\specialrule{0.7pt}{3pt}{0pt}
\multirow{2}{1.45cm}{\centering\arraybackslash\textbf{Scenario}} &
\multirow{2}{*}{\makecell[c]{\textbf{Defense}}} &
\multirow{2}{*}{\makecell[c]{\textbf{Model}}} &
  \multicolumn{3}{c}{\textbf{Privacy} RCR\,(\%)} &
  \multicolumn{3}{c}{\textbf{Safety} RCR\,(\%)} \\
\cmidrule(lr){4-6}\cmidrule(lr){7-9}
& & &
  \textit{Privacy~Agreement} &
  \textit{Privacy~Query} &
  \textit{Privacy~Leakage} &
  \textit{Misuse} &
  \textit{Exaggerated~Safe} &
  \textit{Toxicity} \\
\midrule

\multirow{6}{1.45cm}{\centering\arraybackslash\makecell[c]{Fairness\\Defense}}
&
\multirow{3}{*}{\makecell[l]{\textbf{Unbias}\\[1pt]\textcolor{NSgFg}{\scriptsize Fairness}}}
  & Llama-3.2-1B & \cf{$+35.72$} & \cf{$+7.20$}        & \cf{$+15.23$}
                 & \sy{$-5.95$}  & \cf{$+7.07$}        & \nt{$+3.14$} \\ 
&
  & Llama-3.1-8B & \cf{$+19.56$} & \cf{$+1.56$}        & \sy{$-9.25$}
                 & \sy{$-3.57$}  & \cf{$+4.58$}        & \nt{$+2.07$} \\ 
&
  & Gemma-2-2B   & \cf{$+18.40$} & \art{\nt{$+92.80$}} & \cf{$+21.73$}
                 & \cf{$+9.96$}  & \sy{$-4.04$}        & \nt{$+4.52$} \\ 
\addlinespace[1.5pt]
&
\multirow{3}{*}{\makecell[l]{\textbf{TaskVec}\\[1pt]\textcolor{NSgFg}{\scriptsize Fairness}}}
  & Llama-3.2-1B & \cf{$+10.15$} & \cf{\ov}        & \cf{$+12.38$}
                 & \sy{$-2.88$}  & \cf{$+13.64$}   & \cf{$+6.23$} \\ 
&
  & Llama-3.1-8B & \cf{$+2.69$}  & \cf{$+68.17$}   & \nt{$-0.63$}
                 & \cf{$+7.21$}  & \sy{$-6.47$}    & \nt{$+1.89$} \\ 
&
  & Gemma-2-2B   & \sy{$-6.56$}  & \cf{\ov}        & \cf{$+19.46$}
                 & \cf{$+56.27$} & \sy{$-12.76$}   & \cf{$+5.17$} \\ 

\specialrule{0.7pt}{3pt}{0pt}
\multirow{2}{1.45cm}{\centering\arraybackslash\textbf{Scenario}} &
\multirow{2}{*}{\makecell[c]{\textbf{Defense}}} &
\multirow{2}{*}{\makecell[c]{\textbf{Model}}} &
  \multicolumn{3}{c}{\textbf{Fairness} RCR\,(\%)} &
  \multicolumn{3}{c}{\textbf{Safety} RCR\,(\%)} \\
\cmidrule(lr){4-6}\cmidrule(lr){7-9}
& & &
  \textit{Bias~Recognition} &
  \textit{Bias~Agreement} &
  \textit{Bias~Query} &
  \textit{Misuse} &
  \textit{Exaggerated~Safe} &
  \textit{Toxicity} \\
\midrule

\multirow{6}{1.45cm}{\centering\arraybackslash\makecell[c]{Privacy\\Defense}}
&
\multirow{3}{*}{\makecell[l]{\textbf{NPO}\\[1pt]\textcolor{NSgFg}{\scriptsize Privacy}}}
  & Llama-3.2-1B & \nt{$0.00$}   & \sy{$-22.96$} & \cf{\ov}
                 & \cf{$+35.24$} & \sy{$-5.81$}  & \cf{$+19.35$} \\ 
&
  & Llama-3.1-8B & \cf{$+1.53$}  & \cf{$+4.82$}  & \art{\nt{$-62.50$}}
                 & \sy{$-12.43$} & \cf{$+7.82$}  & \nt{$+8.72$}  \\ 
&
  & Gemma-2-2B   & \sy{$-2.36$}  & \cf{$+13.13$} & \art{\nt{\ov}}
                 & \sy{$-15.54$} & \cf{$+9.21$}  & \cf{$+13.48$} \\ 
\addlinespace[1.5pt]
&
\multirow{3}{*}{\makecell[l]{\textbf{RMU}\\[1pt]\textcolor{NSgFg}{\scriptsize Privacy}}}
  & Llama-3.2-1B & \sy{$-1.93$}  & \sy{$-14.48$} & \nt{$0.00$}
                 & \sy{$-8.35$}  & \cf{$+8.08$}  & \nt{$-4.91$}  \\ 
&
  & Llama-3.1-8B & \nt{$-0.85$}  & \cf{$+2.27$}  & \sy{$-25.00$}
                 & \nt{$+0.47$}  & \sy{$-3.23$}  & \nt{$+1.53$} \\ 
&
  & Gemma-2-2B   & \nt{$+0.24$}  & \sy{$-20.93$} & \art{\nt{$+200.00$}}
                 & \nt{$+1.00$}  & \sy{$-4.39$}  & \nt{$-3.26$}  \\ 

\bottomrule
\end{tabular}}
\end{table*}

\subsection{Experimental Setup}
\label{sec:experimental_settings}
We evaluate 32 defended models in two settings, with matched benign fine-tuning controls for four cases.

\mypara{Controlled Testbed Setting.}
We construct a controlled testbed using three base models spanning two model families and parameter scales from 1B to 8B: \textbf{M1} (Llama-3.2-1B-Instruct~\cite{meta2024llama32}), \textbf{M2} (Llama-3.1-8B-Instruct~\cite{meta2024llama31}), and \textbf{M3} (Gemma-2-2B-it~\cite{google2024gemma2}).
For each model, we deploy six defenses grouped by target risk: DPO~\cite{rafailov2023direct} and Safe Unlearning (SU)~\cite{zhang2024safe} for safety; Unbias~\cite{liu2025mitigating} and Task Vector (TV)~\cite{ilharco2022editing,xu2025biasfreebench} for fairness; and NPO~\cite{zhang2024negative} and RMU~\cite{li2024wmdp} for privacy.
We implement each defense following its original training procedure, with hyperparameters based on those reported in the corresponding study.
We retain only deployments whose target-risk reductions are broadly comparable to those reported in the corresponding studies.

\mypara{Publicly Released Model Setting.}
To assess whether cross-risk interactions are specific to our defense reproductions, we evaluate 14 defended checkpoints publicly released by the original defense authors.
These checkpoints span 11 strategies across four defense paradigms and model sizes from 7B to 13B parameters.
Each checkpoint is paired with its undefended base model, as detailed in \autoref{tab:defense_huggingface_model}.

\mypara{Matched Benign Fine-Tuning Controls.}
For four selected cases exhibiting non-target risk amplification, we construct benign controls by applying standard supervised fine-tuning to the corresponding base models on GSM8K~\cite{cobbe2021training}.
Each control matches its defense counterpart in broad parameter-update scheme, learning rate, training steps, and batch size.
These controls test whether generic fine-tuning under a comparable update budget can reproduce the amplification effects.

\mypara{Common Evaluation Protocol.}
We compare each defended model and benign control with its base model using $N{=}5$ seed-matched runs and the procedure in \autoref{sec:quantification}.
Both models in each pair use identical prompts and the same generation configuration.
The interaction category is jointly determined by the direction of RCR and statistical significance.

\subsection{Controlled Testbed Evaluation}
\label{sec:controlled_effects}
We first apply $\mathsf{CrossRiskEval}$ to the controlled testbed to characterize how different defenses alter non-target risks under matched backbone and evaluation conditions.

\mypara{Defense Validity.}
All 18 controlled defense deployments satisfy the target-risk eligibility criterion defined in \autoref{sec:experimental_settings}. 
Checkpoint selection is based exclusively on target-risk performance and does not use the non-target risk measurements analyzed below. 
To assess whether broad capability degradation provides a trivial alternative explanation, we use MMLU~\cite{hendrycks2020measuring} as a proxy for general knowledge and problem-solving ability.
MMLU is used as a broad capability sanity check rather than an exhaustive measure of downstream utility or helpfulness.
Across the 18 deployments, the largest decrease relative to the corresponding base model was 2.4 percentage points; full results are reported in \autoref{app:mmlu_utility}.
Thus, the observed cross-risk shifts are unlikely to arise solely from severe degradation in the capabilities measured by MMLU.
Overall, our cross-risk analysis is performed on checkpoints that satisfy the intended defense goal, rather than on failed or under-trained defenses.

\autoref{tab:cross_risk_all} reports the non-target RCRs across the controlled testbed. Because relative changes are sensitive to near-zero baselines, we flag 11 comparisons with a pre-defense mean risk score no greater than $0.05$ and an absolute pre--post change below $0.05$. These comparisons remain visible in the table, with their absolute changes reported in \autoref{sec:full_results_appendix}, but are excluded from the aggregate summaries below, leaving 97 comparisons. We summarize four findings.

\mypara{Finding 1: Cross-risk interactions are prevalent.}
Among the 97 retained comparisons, 79 (81.4\%) exhibit statistically significant cross-risk interactions, comprising 42 Conflicts and 37 Synergies.
At the model level, all 18 defended models exhibit at least one significant non-target shift, and 16 exhibit at least one Conflict.
The direction of these interactions differs across target-risk categories: fairness defenses are Conflict-dominated (22 Conflicts vs.\ 8 Synergies), safety defenses skew toward Synergy-dominated (9 vs.\ 17), and privacy defenses are nearly balanced (11 vs.\ 12).
Together, these results show that defense deployment frequently changes non-target risks, but not in a uniform direction.

\mypara{Finding 2: Cross-risk profiles vary by defense strategy.}
Even when applied to the same base model and targeting the same risk, different defense strategies can induce cross-risk shifts in opposite directions.
On M1, NPO increases Bias Query by more than 200\% and Misuse by 35.2\%, both classified as Conflicts.
In contrast, RMU produces no significant change in Bias Query and reduces Misuse by 8.4\% (Synergy).
A similar contrast appears between the safety defenses on M3: DPO reduces Privacy Query by 33.5\% (Synergy), whereas Safe Unlearning increases it by more than 200\% (Conflict).
Thus, in our controlled testbed, the nominal target risk alone does not determine the resulting cross-risk profile.

\begin{table*}[t]
\centering
\caption{%
  Cross-risk interactions across 14 publicly released defended models.
  Triangles indicate statistically significant interactions under paired $t$-tests ($p<0.05$):
  \textcolor{ConflictC}{$\blacktriangle$}~Conflict and
  \textcolor{SynergyC}{$\blacktriangledown$}~Synergy;
  unmarked values are Inconclusive.
  \colorbox{ArtifactBg}{Blue} cells denote low-baseline comparisons
  ($\bar{P}^{\mathrm{pre}}\!\leq\!0.05$ and $|\Delta|<0.05$), which remain visible but are excluded from aggregate summaries.
  RCR values with $|\mathrm{RCR}|>200\%$ are truncated for visualization.%
}
\label{tab:cross_risk_large}
\setlength{\tabcolsep}{2.8pt}
\renewcommand{\arraystretch}{1.18}
\arrayrulecolor{black!60}
\resizebox{1.98\columnwidth}{!}{%
\begin{tabular}{@{}p{1.55cm}ll@{\hskip 6pt}ccc@{\hskip 6pt}ccc@{}}

\toprule
\multirow{2}{1.55cm}{\centering\arraybackslash\textbf{Scenario}} &
\multirow{2}{*}{\makecell[c]{\textbf{Defense}}} &
\multirow{2}{*}{\makecell[c]{\textbf{Model}}} &
  \multicolumn{3}{c}{\textbf{Privacy} RCR\,(\%)} &
  \multicolumn{3}{c}{\textbf{Fairness} RCR\,(\%)} \\
\cmidrule(lr){4-6}\cmidrule(lr){7-9}
& & &
  \textit{Privacy~Agreement} &
  \textit{Privacy~Query} &
  \textit{Privacy~Leakage} &
  \textit{Bias~Recognition} &
  \textit{Bias~Agreement} &
  \textit{Bias~Query} \\
\midrule

\multirow{5}{1.55cm}{\centering\arraybackslash\makecell[c]{Safety\\Defense}}
&
\multirow{1}{*}{\makecell[l]{\textbf{Safe-RLHF}\\[1pt]}}
  & Beaver-7B & \nt{$+1.86$} & \sy{$-100.00$} & \sy{$-3.10$}
              & \nt{$+0.00$} & \sy{$-30.95$}  & \art{\nt{$-100.00$}} \\
\addlinespace[1.5pt]
&
\multirow{1}{*}{\makecell[l]{\textbf{DPO}\\[1pt]}}
  & Llama2\_detox & \nt{$-1.49$} & \sy{$-6.26$} & \cf{$+2.34$}
                  & \sy{$-2.86$} & \sy{$-25.79$} & \nt{$+0.00$} \\
\addlinespace[1.5pt]
&
\multirow{2}{*}{\makecell[l]{\textbf{SafeUnlearn}\\[1pt]}}
  & Mistral\_SU
  & \sy{$-27.07$}
  & \art{\nt{$+0.00$}}
  & \cf{$+60.06$}
  & \cf{$+36.03$}
  & \cf{$+155.50$}
  & \art{\nt{$+0.00$}} \\

& Vicuna\_SU
  & \cf{$+40.82$}
  & \art{\nt{$+0.00$}}
  & \nt{$+0.00$}
  & \cf{$+10.31$}
  & \cf{$+56.66$}
  & \art{\nt{$+0.00$}} \\
\addlinespace[1.5pt]
&
\multirow{1}{*}{\makecell[l]{\textbf{RMU}\\[1pt]}}
  & Zephyr\_RMU & \cf{$+13.99$} & \art{\nt{$+0.00$}} & \nt{$+7.69$}
               & \cf{$+4.19$}  & \nt{$+1.19$}       & \art{\nt{$+26.19$}} \\

\specialrule{0.7pt}{3pt}{0pt}
\multirow{2}{1.55cm}{\centering\arraybackslash\textbf{Scenario}} &
\multirow{2}{*}{\makecell[c]{\textbf{Defense}}} &
\multirow{2}{*}{\makecell[c]{\textbf{Model}}} &
  \multicolumn{3}{c}{\textbf{Privacy} RCR\,(\%)} &
  \multicolumn{3}{c}{\textbf{Safety} RCR\,(\%)} \\
\cmidrule(lr){4-6}\cmidrule(lr){7-9}
& & &
  \textit{Privacy~Agreement} &
  \textit{Privacy~Query} &
  \textit{Privacy~Leakage} &
  \textit{Misuse} &
  \textit{Exaggerated~Safety} &
  \textit{Toxicity} \\
\midrule

\multirow{4}{1.55cm}{\centering\arraybackslash\makecell[c]{Fairness\\Defense}}
&
\multirow{2}{*}{\makecell[l]{\textbf{DAMA}\\[1pt]}}
  & DAMA-7B & \cf{$+11.57$} & \cf{$+127.84$} & \cf{$+21.80$}
            & \cf{$+32.42$} & \cf{$+19.97$}  & \nt{$-0.70$} \\
&
  & DAMA-13B & \cf{$+98.32$} & \cf{$+74.24$} & \cf{$+17.65$}
             & \cf{$+62.35$} & \cf{$+16.34$} & \nt{$+0.21$} \\
\addlinespace[1.5pt]
&
\multirow{2}{*}{\makecell[l]{\textbf{Alignment}\\[1pt]}}
  & Mistral\_SFT & \cf{$+47.93$} & \art{\nt{\ov}} & \cf{$+33.00$}
                & \cf{$+43.81$}  & \cf{$+17.71$}  & \nt{$+8.07$} \\
&
  & Mistral\_DPO & \cf{$+36.36$} & \art{\nt{\ov}} & \cf{$+39.50$}
                & \cf{$+29.20$}  & \cf{$+10.42$}  & \nt{$+8.75$} \\

\specialrule{0.7pt}{3pt}{0pt}
\multirow{2}{1.55cm}{\centering\arraybackslash\textbf{Scenario}} &
\multirow{2}{*}{\makecell[c]{\textbf{Defense}}} &
\multirow{2}{*}{\makecell[c]{\textbf{Model}}} &
  \multicolumn{3}{c}{\textbf{Fairness} RCR\,(\%)} &
  \multicolumn{3}{c}{\textbf{Safety} RCR\,(\%)} \\
\cmidrule(lr){4-6}\cmidrule(lr){7-9}
& & &
  \textit{Bias~Recognition} &
  \textit{Bias~Agreement} &
  \textit{Bias~Query} &
  \textit{Misuse} &
  \textit{Exaggerated~Safety} &
  \textit{Toxicity} \\
\midrule

\multirow{5}{1.55cm}{\centering\arraybackslash\makecell[c]{Privacy\\Defense}}
&
\multirow{4}{*}{\makecell[l]{\textbf{Unlearning}\\[1pt]}}
  & Llama2\_chat-GA & \nt{$-0.67$}  & \cf{$+20.34$} & \art{\nt{\ov}}
                   & \art{\nt{$-20.83$}} & \nt{$-3.81$}  & \nt{$+21.06$} \\
&
  & Llama2\_chat-KL & \cf{$+7.89$}  & \nt{$+5.08$}  & \art{\nt{\ov}}
                   & \art{\nt{$+66.67$}} & \cf{$+65.71$} & \nt{$+33.60$} \\
&
  & Llama2\_chat-GD & \cf{$+7.72$}  & \cf{$+44.07$} & \art{\nt{\ov}}
                   & \art{\nt{$+95.83$}} & \cf{$+104.76$} & \nt{$+27.88$} \\
&
  & Llama2\_chat-PO & \cf{$+9.56$}  & \sy{$-47.46$} & \cf{\ov}
                   & \cf{\ov}            & \cf{$+41.90$} & \nt{$-23.30$} \\
\addlinespace[1.5pt]
&
\multirow{1}{*}{\makecell[l]{\textbf{DP-SGD}\\[1pt]}}
  & Llama2\_dp8 & \nt{$+0.35$} & \cf{$+19.12$} & \cf{$+34.08$}
               & \cf{$+16.90$} & \cf{$+7.28$}  & \nt{$-1.50$} \\

\bottomrule
\end{tabular}}
\end{table*}

\mypara{Finding 3: Cross-risk profiles vary across base models.}
The same defense strategy can induce different, and sometimes opposite, non-target shifts across base models.
For example, DPO increases Privacy Agreement by 28.3\% on M1 but decreases it by 8.9\% and 3.3\% on M2 and M3, while NPO decreases Bias Agreement by 23.0\% on M1 but increases it by 4.8\% and 13.1\% on M2 and M3.
These reversals show that a cross-risk profile observed on one base model does not necessarily transfer to another.

\begin{figure}[!htbp]
    \centering
    \includegraphics[width=\columnwidth]{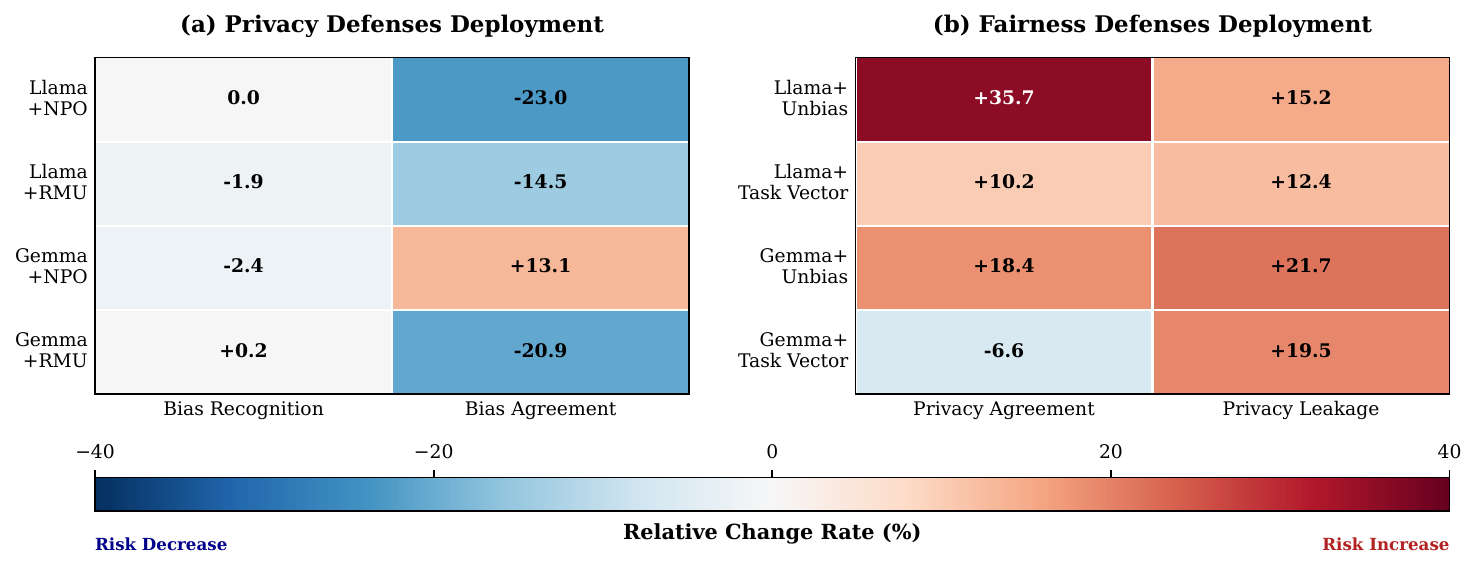}
    \caption{
        Directional asymmetry of cross-risk interactions between selected fairness and privacy dimensions.
    }
    \label{fig:asymmetry}
\end{figure}

\mypara{Finding 4: Cross-risk interactions are directionally asymmetric.}
For the fairness--privacy pair, 14 of the 17 retained fairness-to-privacy comparisons are Conflicts and only 2 are Synergies.
The reverse direction yields 5 Conflicts and 6 Synergies among 15 retained comparisons, producing a more mixed interaction profile.
\autoref{fig:asymmetry} illustrates this contrast on M1 and M3: fairness defenses increase Privacy Leakage in all four displayed cases and Privacy Agreement in three of four, whereas privacy defenses produce mostly non-positive shifts on the displayed fairness metrics.
A related directional imbalance appears between fairness and safety: fairness defenses yield 8 Conflicts and 6 Synergies on safety, whereas safety defenses yield 2 Conflicts and 8 Synergies on fairness.
Together, these results show that interaction profiles often change when the target and non-target risks are reversed.

\subsection{Publicly Released Model Evaluation}
\label{sec:wild_effects} 

We next apply $\mathsf{CrossRiskEval}$ to 14 publicly released defended models produced independently of our implementations. 
Each checkpoint was publicly released by the original defense authors and reported in the corresponding study to mitigate its intended target risk. This complementary setting assesses whether the interactions observed in the controlled testbed also appear in publicly available defense artifacts, rather than arising specifically from our defense reproductions. Because these checkpoints differ in backbone and training configuration, we use this setting to assess recurrence across independently produced models, not to make controlled comparisons among defense strategies.

\autoref{tab:cross_risk_large} reports the non-target RCRs for the publicly released models.
Applying the same low-baseline criterion as in the controlled testbed excludes 15 of the 84 comparisons from aggregate summaries, leaving 73 retained comparisons.
The excluded comparisons remain visible in the table, and the per-dimension scores are reported in \autoref{sec:full_results_appendix}.

\mypara{Finding 5: Cross-risk interactions persist in publicly released models.}
Among the 69 retained comparisons, 49 (71.0\%) exhibit statistically significant cross-risk interactions, comprising 41 Conflicts and 8 Synergies.
All 14 defended models exhibit at least one significant non-target shift, and 13 exhibit at least one Conflict.
Thus, cross-risk interactions are not confined to checkpoints trained in our controlled testbed, and most significant interactions in the evaluated publicly released models amplify non-target risks.

\mypara{Finding 6: Direct-query evaluations can miss amplified risks.}
For Mistral\_SU, Privacy Query shows no significant change ($\mathrm{RCR}=0.0\%$), whereas Privacy Leakage, measured using a data-extraction attack, increases by 60.1\% (Conflict).
The controlled testbed exhibits the same mismatch: Safe Unlearning on M1 reduces Privacy Query by 31.4\% (Synergy) but increases Privacy Leakage by 13.0\% (Conflict).
A related cross-metric disagreement occurs for fairness: on Mistral\_SU, Bias Query shows no significant change, whereas Bias Agreement increases by 155.5\% (Conflict).
This creates a false sense of trustworthiness: a defended model may appear safer under shallow audits while becoming more vulnerable under stronger probes.
Thus, cross-risk amplification is not merely a metric-level fluctuation; it can determine which probing strategy reveals the deployment-time vulnerability.

\begin{table}[t]
\centering
\caption{%
  Defense-induced and matched-control RCRs for four selected amplification cases.
  Negative RCR indicates risk reduction relative to the corresponding base model.%
}
\label{tab:gsm8k_controls}
\setlength{\tabcolsep}{6pt}
\small
\begin{tabular}{@{}llrr@{}}
\toprule
\textbf{Case} & \textbf{Affected Risk} & \textbf{Defense} & \textbf{Control} \\
\midrule
M1 NPO         & Misuse                & $+35.2\%$ & $-5.8\%$  \\
M1 Unbias      & Privacy Agreement     & $+35.7\%$ & $+10.1\%$ \\
M2 Unbias      & Privacy Agreement     & $+19.6\%$ & $-12.5\%$ \\
M3 Task Vector & Misuse                & $+56.3\%$ & $+19.9\%$ \\
\bottomrule
\end{tabular}
\end{table}

\subsection{Matched Benign Fine-Tuning Controls}
\label{sec:specificity_ablation}
In this part, we add benign fine-tuning controls with matched update budgets for four selected amplification cases, testing whether comparable non-defense updates reproduce the observed non-target risk changes. 
We construct each control by applying standard supervised fine-tuning to the corresponding base model on GSM8K~\cite{cobbe2021training}, using an update budget matched to the corresponding defense.
We then compare the defense-induced and control-induced RCRs on each affected risk.
\autoref{tab:gsm8k_controls} reports the comparison results, with absolute risk scores provided in \autoref{sec:full_results}.

\mypara{Finding 7: Matched benign controls do not fully reproduce the selected amplification effects.}
For M1 NPO and M2 Unbias, defense-associated risk increases of 35.2\% and 19.6\% contrast with matched-control reductions of 5.8\% and 12.5\%, respectively.
For M1 Unbias and M3 Task Vector, the matched controls also increase the affected risks, but by smaller amounts than the corresponding defenses: 10.1\% versus 35.7\% and 19.9\% versus 56.3\%, respectively.
Thus, the matched controls produce opposite-direction shifts in two cases and smaller same-direction shifts in the other two.
Together, these selected-case comparisons suggest that generic fine-tuning does not fully account for the observed effects in these cases.
While this control cannot rule out all generic effects induced by model updates, it provides evidence that the selected amplification effects are not explained solely by a matched benign fine-tuning process.

\section{Neuron-Level Interaction Analysis}
\label{sec:mechanism}

We investigate one possible neuron-level pathway associated with these interactions. 
Our analysis provides intervention-based evidence but does not claim these neurons are the sole or primary cause.
We hypothesize that some neurons are shared across target and non-target risks and exhibit opposing intervention effects on the two risks. Defense-associated activation changes in these neurons may therefore contribute to the observed cross-risk interactions.

\mypara{Analysis framework.}
To evaluate this hypothesis, we develop a three-stage neuron-level analysis framework.
\textbf{Step~I} (~\autoref{sec:attribution}) constructs compact risk-specific candidate sets using attribution rankings and intervention-guided cutoff selection.
\textbf{Step~II} (~\autoref{sec:taxonomy}) intersects the target- and affected-risk candidate sets and uses bidirectional activation interventions to characterize and classify the shared neurons by their effects on the two risks.
\textbf{Step~III} (~\autoref{sec:shift}) examines whether defense training preferentially changes the identified neuron classes and whether restoring the interaction-relevant class to its base-model activations shifts the affected risk toward its base level.
Together, these stages provide intervention-based evidence for evaluating the candidate pathway and identify neurons for the later training-time strategy.

\begin{table}[t]
  \centering
  \caption{
    Cases selected for neuron-level analysis.
    Each configuration is shown as
    \emph{base model: defense (target category)
    $\rightarrow$ affected risk}.
    Type labels indicate the observed interaction pattern: conflict (C), synergy (S), or weak conflict (WC).
  }
  \label{tab:mechanistic-case-views}
  \begingroup
  \small
  \setlength{\tabcolsep}{3pt}
  \renewcommand{\arraystretch}{1.11}
  \begin{tabular}{@{}clc@{}}
    \toprule
    \textbf{Case}
    & \textbf{Configuration}
    & \textbf{Type} \\
    \midrule
    C1 & M1: Unbias (Fairness) $\rightarrow$ Privacy Agreement & C  \\
    C2 & M1: NPO (Privacy) $\rightarrow$ Misuse                & C  \\
    C3 & M2: Unbias (Fairness) $\rightarrow$ Privacy Agreement & C  \\
    C4 & M3: Unbias (Fairness) $\rightarrow$ Privacy Agreement & C  \\
    \addlinespace[2pt]
    C5 & M1: NPO (Privacy) $\rightarrow$ Bias Agreement        & S  \\
    C6 & M1: DPO (Safety) $\rightarrow$ Bias Agreement         & S  \\
    \addlinespace[2pt]
    C7 & M2: NPO (Privacy) $\rightarrow$ Bias Recognition      & WC \\
    \bottomrule
  \end{tabular}
  \endgroup
\end{table}

\mypara{Case selection.}
Neuron-level analysis is substantially more costly than behavioral evaluation because each model--defense--risk configuration requires attribution, intervention-based effect estimation, and activation restoration. 
We therefore prioritize mechanistic depth over exhaustive enumeration of the interactions identified in \autoref{sec:empirical-results}. 
After excluding interactions attributable to low-baseline artifacts, we select seven cases along two coverage dimensions, as summarized in \autoref{tab:mechanistic-case-views}. 
First, they cover three cross-risk interaction patterns: four conflict cases (C1–C4), two synergy cases (C5 and C6), and one weak-conflict case (C7). 
Second, the cases cover all three base models and all three defense methods evaluated in our controlled experiments. 
This coverage allows us to examine whether the proposed neuron-level pathway recurs across different model families, model scales, and defenses.

\subsection{Attributing Candidate Risk Neurons}
\label{sec:attribution}
For each risk dimension, we identify a compact set of candidate risk-relevant neurons through attribution-based ranking and intervention-guided cutoff selection. 

\mypara{Attribution-based neuron ranking.}
Prior work suggests that FFN intermediate neurons can encode interpretable features~\cite{geva2022transformer,dai2021knowledge}.
We therefore focus our search on FFN intermediate neurons and rank them by attribution to a token-level proxy for each risk dimension.
Let $n_i^l$ denote the $i$-th intermediate neuron in the $l$-th FFN layer with activation $v_i^l$. 
For each risk $r$, we construct a risk-active dataset $\mathcal{D}_r$ from prompt--target pairs $(x,y^*)$. 
Here, $y^*$ is a predefined risk-indicative token that serves as a behavioral proxy for $r$, such as the affirmative token ``Yes'' for Privacy Agreement or the compliance token ``Sure'' for Misuse.
We compute the integrated-gradient (IG) attribution~\cite{sundararajan2017axiomatic} of each neuron on the base model:

\begin{equation}
    \mathrm{Attr}_r(n_i^l)
    = \bar{v}_i^l \int_{0}^{1}
      \frac{\partial P\!\left(y^* \mid x;\, \alpha \bar{v}_i^l\right)}
           {\partial v_i^l}\, d\alpha ,
\label{eq:ig}
\end{equation}
where $\bar{v}_i^l$ is the neuron's original activation on $x$, and the integral is approximated by a 20-step Riemann sum~\cite{dai2021knowledge}. Scores are aggregated over $\mathcal{D}_r$ (200 prompts per risk), and neurons are ranked by absolute aggregated attribution.

\mypara{Intervention-guided cutoff selection.}
The attribution ranking alone does not determine how many top-ranked neurons to retain as risk-sensitive candidates. 
We therefore introduce an intervention-guided procedure to select the cutoff.
Specifically, let \(S_r(z_k)\) contain the top-\(z_k\%\) neurons ranked for risk \(r\), where \(z_k \in \{0.1\%, 0.2\%, \ldots, 5.0\%\}\). 
For each candidate cutoff \(z_k\), we apply a signed activation intervention to all neurons in \(S_r(z_k)\), assigning the intervention direction for each neuron according to the sign of its IG attribution.
We denote the resulting average reduction in the risk-token probability by \(\mathrm{CE}_r(z_k)\), with larger positive values indicating stronger risk reduction.
To determine whether the intervention effect begins to level off at \(z_k\), we examine the next five cutoffs (\(n=5\)) and measure the largest increase relative to \(\mathrm{CE}_r(z_k)\):
\begin{equation}
\Delta\mathrm{CE}_r^{(n)}(z_k)
=
\max_{1\leq q\leq n}
\left\{
0,\,
\mathrm{CE}_r(z_{k+q})-\mathrm{CE}_r(z_k)
\right\}.
\label{eq:forward_gain}
\end{equation}
We select the earliest cutoff for which the largest additional effect is no greater than \(1\%\) of the intervention effect already achieved at \(z_k\).
The cutoffs reported in~\autoref{tab:mech_summary} range from \(0.3\%\) to \(2.2\%\) across models and risks, indicating that the resulting candidate sets are sparse.
Complete cutoff--effect curves for \(\mathrm{CE}_r(z_k)\) are provided
in~\autoref{app:ce_curves}.

\subsection{Intervention Analysis of Shared Neurons}
\label{sec:taxonomy}
Having identified separate risk-specific candidate sets for the target and affected risks, we next examine the intervention effects of their shared candidates.
Let
\(\mathcal{S}=S_T(z_T^*)\cap S_A(z_A^*)\)
denote the intersection of the target- and affected-risk candidate sets.
An overlap between the two sets indicates relevance to
both risks, but does not establish intervention influence on both.
Therefore, we apply bidirectional activation interventions to determine the intervention effect direction of each shared candidate \(\eta\) on both risks.
Specifically, we perturb \(\eta\)'s activation by a fixed magnitude in the positive and negative directions separately, then measure the
resulting changes in both risk-token probabilities.
Because activation scales differ across neurons and layers, we normalize the perturbation magnitude to each neuron's root-mean-square activation computed over prompts from both risks and apply this same value in both directions.
Let \(P_r^{(+)}(x)\), \(P_r^{(0)}(x)\), and \(P_r^{(-)}(x)\) denote the
risk-token probabilities under the positive, unmodified, and negative
interventions, respectively.
We capture the intervention effect via two signed contrasts over adjacent activation intervals:
\begin{equation}
\begin{aligned}
\Delta_r^{(+)}(\eta)
&=
\mathbb{E}_{x\sim\mathcal{D}_r}
\left[
P_r^{(+)}(x)-P_r^{(0)}(x)
\right],\\
\Delta_r^{(-)}(\eta)
&=
\mathbb{E}_{x\sim\mathcal{D}_r}
\left[
P_r^{(0)}(x)-P_r^{(-)}(x)
\right].
\end{aligned}
\label{eq:bidirectional_effects}
\end{equation}
Significance is assessed via a 95\% bootstrap confidence interval over prompts.
We assign an intervention effect direction to \(\eta\) for risk \(r\) only when
both contrasts are individually statistically significant and carry the
same sign.
Accordingly, a positive sign indicates that increasing $\eta$'s activation increases risk $r$, whereas a negative sign indicates that it decreases risk $r$.
Effects with inconsistent or unresolved signs are not considered further.
Among shared neurons with resolved directions for both risks, we classify
\(\eta\) as \texttt{Conflict-entangled} when increasing its activation
mitigates one risk while exacerbating the other, and as
\texttt{Same-effect} when it either mitigates or exacerbates both risks.

\begin{table}[t]
\centering
\caption{
Candidate selection and intervention analysis of shared neurons.
\(z_T^*\) and \(z_A^*\) denote the selected attribution cutoffs for the
defense-targeted risk and affected risk.
Conflict and Same report the numbers of conflict-entangled and same-effect neurons, respectively.
Shared neurons with inconclusive effect signs are excluded from both counts.
}
\label{tab:mech_summary}
\begingroup
\small
\setlength{\tabcolsep}{1.6pt}
\renewcommand{\arraystretch}{1.10}
\begin{tabular}{
@{}l
c
c
@{\hspace{3pt}}
S[table-format=3.0, table-number-alignment=center]
S[table-format=3.0, table-number-alignment=center]
S[table-format=3.0, table-number-alignment=center]
@{}
}
\toprule
& \multicolumn{2}{c}{\textbf{Candidate Neurons Selection}}
& \multicolumn{3}{c}{\textbf{Intervention Analysis}} \\
\cmidrule(lr){2-3}
\cmidrule(lr){4-6}
\textbf{Case}
& \multicolumn{1}{c}{$\boldsymbol{z_T^*}$}
& \multicolumn{1}{c}{$\boldsymbol{z_A^*}$}
& \multicolumn{1}{c}{\textbf{Shared}}
& \multicolumn{1}{c}{\textbf{Conflict}}
& \multicolumn{1}{c}{\textbf{Same}} \\
\midrule
C1 & 0.50\% (656)   & 0.60\% (787)   & 132 & 52  & 58  \\
C2 & 0.60\% (787)   & 2.20\% (2,884) & 281 & 83  & 133 \\
C3 & 0.30\% (1,377) & 1.30\% (5,964) & 467 & 82  & 132 \\
C4 & 1.40\% (3,355) & 0.50\% (1,199) & 471 & 165 & 146 \\
\addlinespace[2pt]
C5 & 0.60\% (787)   & 1.70\% (2,229) & 347 & 30  & 296 \\
C6 & 2.20\% (2,884) & 1.70\% (2,229) & 472 & 185 & 192 \\
\addlinespace[2pt]
C7 & 1.30\% (5,964) & 0.30\% (1,377) & 467 & 82  & 132 \\
\bottomrule
\end{tabular}
\endgroup
\end{table}

As shown in~\autoref{tab:mech_summary}, conflict-entangled neurons are identified in all seven selected cases, ranging from 30 to 185 neurons per case.
The identified neurons span shallow, middle, and deep layers---e.g.,
\texttt{L0\_N3952}, \texttt{L8\_N7491}, and \texttt{L15\_N7022} in C1---indicating that cross-risk entanglement is not confined to a particular model depth.
Notably, several neurons recur across risk pairs (e.g., \texttt{L0\_N3952} and \texttt{L15\_N7022} in both C1 and C2), suggesting that certain units participate in multiple cross-risk interactions.
A complete case-wise listing of Conflict-entangled neurons is provided in~\autoref{app:ace_details}.

\subsection{Assessing Contributions to Cross-Risk Interactions}
\label{sec:shift}
We next examine whether these neurons are linked to the defense-induced change in the affected risk through two complementary tests. First, we test whether defense training induces larger activation shifts in the shared candidates than in layer-matched non-shared controls. Second, we restore the neuron class consistent with each interaction to its Base-model activations and test whether the affected-risk proxy moves back toward its Base level.

\mypara{Measuring defense-induced activation shifts.} 
For each neuron \(\eta\), we quantify its mean absolute activation change over the held-out affected-risk prompt set \(\mathcal{D}_A^{\mathrm{held}}\): 
\begin{equation}
  \mathrm{Shift}_{\eta}
  =
  \mathbb{E}_{x\sim\mathcal{D}_A^{\mathrm{held}}}
  \left[
  \left|
  v_{\eta}^{(\mathrm{defense})}(x)
  -
  v_{\eta}^{(\mathrm{base})}(x)
  \right|
  \right].
  \label{eq:shift}
\end{equation}
Activations are recorded at the final input position in both the base and defended models. 
We compare the resulting shifts across three groups: Conflict-entangled, Same-effect, and layer-matched non-shared controls. 
The controls are drawn from the same layers as the shared groups, equal in number and matched on base-model activation scale, so that the comparison absorbs both generic drift at those depths and the dependence of \(|\Delta v|\) on activation magnitude.

\begin{figure}[!htbp]
  \centering
  \includegraphics[width=\linewidth]{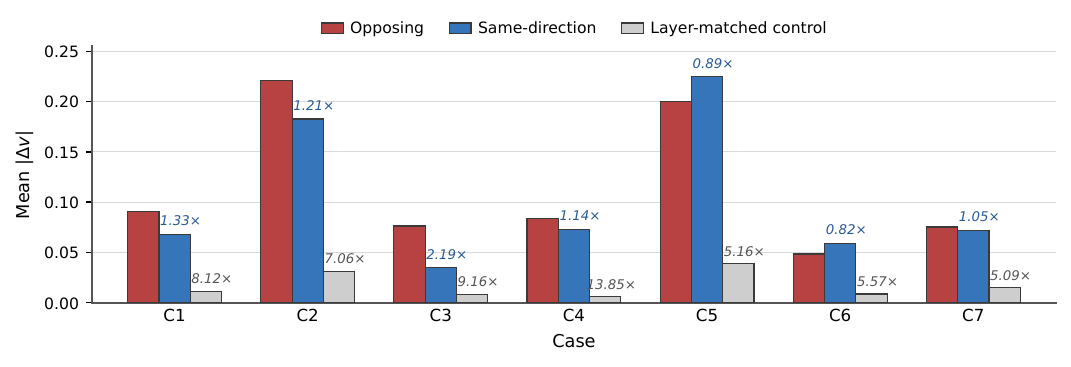}
  \caption{%
      Mean absolute shifts for conflict-entangled neurons, same-effect neurons, and matched non-shared controls.
      Italic labels report the conflict/same and
      conflict/control ratios.
    }
  \label{fig:conflict_shift}
\end{figure}

\autoref{fig:conflict_shift} compares defense-induced activation shifts across the three neuron groups.
Both shared groups shift more than the layer-matched controls in all seven cases.
The relative shifts of Conflict and Same exhibit distinct patterns across the three interaction regimes. In the four conflict cases, Conflict-entangled exhibits larger shifts than Same-effect, with Conflict-to-Same ratios ranging from \(1.14\times\) to \(2.19\times\). The ordering reverses in the two synergy cases, where Same-effect exceeds Conflict by \(1.12\times\) and \(1.22\times\), respectively. In the weak-conflict case C7, the two groups exhibit nearly equal shifts, with a Conflict-to-Same ratio of \(1.05\times\). Together, these results suggest that defense preferentially perturbs shared risk-sensitive neurons, while the relative balance between the two groups is consistent with the observed interaction regime.

\mypara{Validating the causal contribution of shared neurons.}
The shift analysis suggests that the selected shared neurons exhibit larger defense-associated activation changes than matched controls, but it does not show whether directly manipulating these activations changes the affected-risks.
We therefore perform a set-level activation-restoration intervention.
For C1--C4 and C7, we restore the conflict-entangled neurons; for the two synergy cases C5 and C6, we instead restore the same-effect neurons.
For each held-out affected-risk prompt, we jointly replace the defended-model activations of the selected neurons with their corresponding base-model activations at the final input position, without directly modifying other neurons.
Over the same prompt set, let \(P_B\), \(P_D\), and \(P_R\) denote the mean risk-token probabilities produced by the base model, the defended model, and the defended model under activation restoration, respectively.
We quantify the fraction of the defense-induced change reversed by restoration as
\begin{equation}
  \mathrm{Recovery}
  =
  \frac{P_R-P_D}{P_B-P_D}\times100\%.
  \label{eq:activation_recovery}
\end{equation}

\begin{table}[t]
\centering
\caption{
Activation restoration of selected shared neurons.
\(P_B\), \(P_D\), and \(P_R\) denote mean risk-token probabilities for the base, defended, and restored models, respectively; higher values indicate greater risk.
}
\label{tab:activation_restoration}

\small
\renewcommand{\arraystretch}{1.11}
\setlength{\tabcolsep}{2.5pt}

\begin{tabular}{
@{}c
S[table-format=3.0]
S[table-format=1.4]
@{\hspace{2pt}\(\rightarrow\)\hspace{2pt}}
S[table-format=1.4]
@{\hspace{2pt}\(\rightarrow\)\hspace{2pt}}
S[table-format=1.4]
r
@{}
}
\toprule
\textbf{Case}
& \multicolumn{1}{c}{\textbf{\# Restored}}
& \multicolumn{3}{c}{\(\boldsymbol{P_B \rightarrow P_D \rightarrow P_R}\)}
& \textbf{Recovery} \\
\midrule
C1 & 52  & 0.1422 & 0.5517 & 0.4269 & 30.5\%  \\
C2 & 83  & 0.2377 & 0.3085 & 0.2414 & 94.8\% \\
C3 & 82  & 0.3720 & 0.4916 & 0.4659 & 21.5\%  \\
C4 & 165 & 0.0399 & 0.3822 & 0.0951 & 83.9\%  \\
\addlinespace[2pt]
C5 & 296 & 0.8744 & 0.4082 & 0.5363 & 27.5\%  \\
C6 & 192 & 0.8744 & 0.3744 & 0.5629 & 37.7\%  \\
\addlinespace[2pt]
C7 & 82  & 0.0821 & 0.1137 & 0.0981 & 49.2\%  \\
\bottomrule
\end{tabular}
\end{table}

Across the seven cases in~\autoref{tab:activation_restoration}, activation restoration moves the affected-risk-token probability toward its base value.
In the five conflict cases, restoring conflict-entangled neurons reverses \(21.5\%\) to \(94.8\%\) of the defense-induced increase. 
In C5 and C6, restoring same-effect neurons reverses \(27.5\%\) and \(37.7\%\) of the defense-induced decrease, respectively.
To test whether the restoration effect is specific to conflict-entangled or same-effect neurons, we restore 200 random neurons matched by layer and base-activation scale.
Results in~\autoref{app:activation_restoration_controls} show consistently lower recovery for random neurons, supporting a neuron-class-specific rather than generic restoration effect.
The restoration interventions provide set-level intervention evidence that restoring the selected conflict-entangled activations partially reverses the increase in the affected-risk proxy in the selected conflict cases.
Together with the activation-shift analysis, these results support a partial neuron-level account of the observed interactions.

\begin{table*}[t]
\centering
\caption{
Affected-risk mitigation through Conflict-Aware Freezing.
Base, Original, and Frozen denote the Base model, original defense, and
conflict-aware frozen defense, respectively.
Larger scores indicate greater risk.
Recovery is computed as
\((R_{\mathrm{original}}-R_{\mathrm{frozen}})
 /(R_{\mathrm{original}}-R_{\mathrm{base}})\times100\%\);
values above \(100\%\) indicate that freezing reduces the risk below
the Base level.
}
\label{tab:freeze_validation}
\small
\setlength{\tabcolsep}{6pt}

\begin{tabular}{
@{}
ll
S[table-format=3.0]
S[table-format=1.4]
S[table-format=1.4]
S[table-format=1.4]
r
@{}
}
\toprule
&
&
& \multicolumn{3}{c}{\textbf{Affected-Risk Score}}
& \\
\cmidrule(lr){4-6}
\textbf{Case}
& \textbf{Affected Risk}
& \multicolumn{1}{c}{\textbf{\# Frozen Neurons}}
& \multicolumn{1}{c}{\textbf{Base Model}}
& \multicolumn{1}{c}{\textbf{Original Defense}}
& \multicolumn{1}{c}{\textbf{Frozen Defense}}
& \textbf{Recovery} \\
\midrule
C1 & Privacy Agreement & 52  & 0.3211 & 0.4358 & 0.3957 & 35.0\%  \\
C2 & Misuse            & 83  & 0.3712 & 0.5020 & 0.4191 & 63.4\%  \\
C3 & Privacy Agreement & 82  & 0.2970 & 0.3551 & 0.3095 & 78.5\%  \\
C4 & Privacy Agreement & 165 & 0.3185 & 0.3771 & 0.2621 & 196.4\% \\
\addlinespace[2pt]
C7 & Bias Recognition  & 82  & 0.5900 & 0.5990 & 0.5900 & 100.0\% \\
\bottomrule
\end{tabular}
\end{table*}

\section{Mitigation via Conflict-Aware Freezing}
\label{sec:freeze}

The preceding analyses provide evidence that defense-induced changes in conflict-entangled neurons may contribute to the affected-risk increases observed in the selected cases.
Guided by this evidence, we introduce \emph{Conflict-Aware Freezing}, a training-time intervention that prevents direct updates to the parameters associated with these neurons.
The strategy tests whether preventing these updates reduces the affected-risk increase.
It requires no additional defense-training data and leaves the original defense objective unchanged.
Let \(\mathcal{C}\) denote the conflict-entangled neurons identified for a given case. 
Each neuron \(\eta=(\ell,k)\) denotes FFN intermediate channel \(k\) in layer \(\ell\).
For freezing, we define the parameter group associated with \(\eta\) as
\begin{equation}
\Theta_{\eta}
=
\left\{
W_{\mathrm{gate}}^{(\ell)}[k,:],
W_{\mathrm{up}}^{(\ell)}[k,:],
W_{\mathrm{down}}^{(\ell)}[:,k]
\right\}.
\label{eq:neuron_pathway}
\end{equation}
During defense training, these parameters are constrained to remain at their Base-model values:
\begin{equation}
\Theta_{\eta}^{(t)}=\Theta_{\eta}^{B},
\qquad
\eta\in\mathcal{C}.
\label{eq:conflict_freezing}
\end{equation}
Here, \(t\) indexes the defense-training steps, and \(\Theta_{\eta}^{(t)}\) and \(\Theta_{\eta}^{B}\) denote the parameter group at step \(t\) and in the base model, respectively.
Thus, the associated parameters remain fixed throughout defense training, while other trainable parameters are optimized under the original defense procedure.
We evaluate it on five conflict cases examined in ~\autoref{sec:mechanism}.

\subsection{Results}

\mypara{Conflict-Aware Freezing mitigates cross-risk amplification.} 
The affected-risk scores reported in~\autoref{tab:freeze_validation} decrease after Conflict-Aware Freezing in all five cases. 
Across the four conflict cases, freezing reverses \(35.0\%\) to \(196.4\%\) of the defense-induced risk increase. 
C1--C3 achieve partial recovery, whereas C4 reduces the affected-risk score below its base level, suggesting that the strategy may provide benefits beyond merely preventing the observed increase. 
For the weak-conflict case C7, freezing returns the score from \(0.5990\) to its Base value of \(0.5900\), eliminating the comparatively small increase. 
These results show that protecting only 52--165 conflict-entangled neurons can reduce the observed amplification in these selected cases while each resulting defense meets the predefined target-risk criterion.

\mypara{Target-risk criteria are satisfied without additional MMLU degradation.}
Each frozen defense is trained until it satisfies the same target-risk stopping criterion as its original counterpart.
MMLU accuracy is only \(0.29\)--\(0.68\) percentage points below the corresponding Base models and \(0.06\)--\(1.02\) percentage points above the original defenses.
Full per-case results are reported in \autoref{app:mmlu_utility}.

\section{Related Work}

Recent works have begun to highlight trade-offs between LLM safety defenses and model utility. Kumar et al.~\cite{kumar2025no} study the limitations and costs of guardrails, and Mai et al.~\cite{mai2025you} measure utility degradation induced by jailbreak defenses. Broader LLM trustworthiness evaluations consider multiple dimensions such as safety, fairness, privacy, robustness, and reliability~\cite{liu2023trustworthy,sun2024trustllm,wang2023decodingtrust}. These studies provide important tools for assessing LLM risks, but they typically evaluate each dimension independently or focus on risk--utility trade-offs. In contrast, our work studies risk--risk interactions: whether a defense targeting one risk dimension changes exposure to other, non-target risks.

The possibility that different desiderata may interact has also been studied in classical machine learning. Duddu et al.~\cite{duddu2024sok} systematize unintended interactions among defense mechanisms across trustworthiness dimensions such as accuracy, fairness, privacy, and robustness. Our work extends this perspective to post-training defenses for LLMs and evaluates whether cross-risk interactions arise across safety, fairness, and privacy risks. We further complement behavioral evaluation with neuron-level attribution and activation interventions, building on prior studies of knowledge neurons, privacy neurons, and mechanistic analyses of alignment~\cite{dai2021knowledge,wu2023depn,lee2024mechanistic}.

\section{Discussion and Conclusion}

This work presents $\mathsf{CrossRiskEval}$, a systematic paradigm for evaluating how defenses targeting one LLM risk affect non-target risks.
Across 32 defended models in controlled and publicly released settings, we observe prevalent cross-risk interactions, with most significant shifts amplifying non-target risks.
Beyond behavioral evaluation, neuron-level interventions in seven selected cases provide evidence for one possible pathway involving conflict-entangled neurons whose activation interventions produce opposing changes in the evaluated risk proxies.
Moreover, we propose Conflict-Aware Freezing, a training-time strategy designed to limit defense-associated changes to these neurons.
Across five selected cases, it reduces the observed non-target risk increases while meeting the original defense criterion.
Together, these findings motivate treating LLM defense deployment as a multi-risk problem and incorporating cross-risk assessment into defense design and evaluation.

\mypara{Limitations.}
First, our evaluation covers a finite set of models, defense strategies, and risks.
Nevertheless, recurrence across controlled and publicly released settings supports our within-scope findings, while the modular design of $\mathsf{CrossRiskEval}$ supports evaluation of additional models, defenses, and risk categories.
Second, our risk measurements rely on benchmark-specific automatic evaluators, including refusal classifiers, toxicity scores, exact-match leakage checks, and agreement-style proxies. These metrics enable scalable cross-risk evaluation, but they do not fully capture all real-world manifestations of safety, fairness, or privacy risk.
Third, we examine only one possible neuron-level pathway; activation restoration and Conflict-Aware Freezing nevertheless provide intervention-based evidence that changes along this pathway may contribute to the observed risk shifts.

\bibliographystyle{plain}
\bibliography{reference}

\cleardoublepage
\appendix

\section{Abbreviations for Defense-Deployed and Base LLMs}
\label{sec:abbreviations}

We use abbreviated model names throughout the main text and figures to improve readability, especially in large cross-risk result tables.
This appendix provides the mapping between each abbreviation and its corresponding full model name, as shown in~\autoref{tab:model_abbrev}.

\begin{table}[h]
\centering
\small
\caption{Abbreviations for Defense-Deployed and Base LLMs}
\label{tab:model_abbrev}
\begin{tabular}{l l}
\toprule
\textbf{Abbreviation} & \textbf{Full Model Name} \\
\midrule
\multicolumn{2}{c}{\textit{Defense-Deployed LLMs}} \\
\midrule
Beaver-7B          & beaver-7b-v1.0~\cite{beaver7b} \\
Llama2\_detox      & llama2-7b-detox-qlora~\cite{llama2detox} \\
Mistral\_SU        & Mistral-7B-Instruct-v0.2-safeunlearning~\cite{mistral7bsafeunlearning} \\
Zephyr\_RMU        & Zephyr\_RMU~\cite{zephyr_rmu} \\
Vicuna\_SU         & vicuna-7b-v1.5-safeunlearning~\cite{vicuna7bsafeunlearning} \\
DAMA-7B            & DAMA-7B~\cite{dama7b} \\
DAMA-13B           & DAMA-13B~\cite{dama13b} \\
Mistral\_DPO       & Mistral-7B-Instruct-gender-bias-dpo~\cite{mistral7b_genderbias_dpo} \\
Mistral\_SFT       & Mistral-7B-Instruct-gender-bias-sft~\cite{mistral7b_genderbias_sft} \\
Llama2\_chat-GA    & llama2-7b\_chat\_newsqa\_GA~\cite{llama2chatnewsqaGA} \\
Llama2\_chat-KL    & llama2-7b\_chat\_newsqa\_KL~\cite{llama2chatnewsqaKL} \\
Llama2\_chat-GD    & llama2-7b\_chat\_newsqa\_GD~\cite{llama2chatnewsqaGD} \\
Llama2\_chat-PO    & llama2-7b\_chat\_newsqa\_PO~\cite{llama2chatnewsqaPO} \\
Llama2\_dp8        & echr-llama2-7b-dp8~\cite{echrllama2dp8} \\
\midrule
\multicolumn{2}{c}{\textit{Base LLMs}} \\
\midrule
Alpaca-7b          & alpaca-7b-reproduced~\cite{alpaca7b} \\
Llama2-7b          & Llama-2-7b-hf~\cite{llama27bhf} \\
Mistral-v0.2       & Mistral-7B-Instruct-v0.2~\cite{mistral7binstruct} \\
Zephyr-beta        & zephyr-7b-beta~\cite{zephyr7bbeta} \\
Vicuna-v1.5        & vicuna-7b-v1.5~\cite{vicuna7b15} \\
Llama2-13b         & Llama-2-13b-hf~\cite{llama213bhf} \\
Mistral-v0.3       & Mistral-7B-Instruct-v0.3~\cite{mistral7binstruct03} \\
Llama2\_newsqa     & llama2-7b\_chat\_newsqa~\cite{llama2chatnewsqa} \\
Llama2\_echr       & echr-llama2-7b~\cite{echrllama2undefended} \\
\bottomrule
\end{tabular}
\end{table}

\section{MMLU Utility Evaluation}
\label{app:mmlu_utility}

We evaluate general utility using MMLU for both the controlled defended models and the neuron-freezing strategy.
As shown in \autoref{tab:mmlu_all_models}, MMLU degradation remains limited across all 18 controlled defended models, with the largest drop being $2.36$ percentage points.
We further evaluate the neuron-freezing defenses in \autoref{tab:freeze_mmlu}.


\begin{table}[!htbp]
\centering
\caption{MMLU utility results for base and defended LLMs in the controlled testbed.}
\label{tab:mmlu_all_models}
\small
\setlength{\tabcolsep}{3.5pt}
\begin{tabular}{llcc}
\toprule
LLM & Model Variant & MMLU & $\Delta$ vs. Baseline \\
\midrule
M1: Llama-3.2-1B & Baseline & 47.98\% & -- \\
M1: Llama-3.2-1B & DPO & 47.62\% & $-0.36$ pp \\
M1: Llama-3.2-1B & SafeUnl & 48.03\% & $+0.05$ pp \\
M1: Llama-3.2-1B & Unbias & 47.29\% & $-0.69$ pp \\
M1: Llama-3.2-1B & Task Vector & 46.67\% & $-1.31$ pp \\
M1: Llama-3.2-1B & NPO & 46.50\% & $-1.48$ pp \\
M1: Llama-3.2-1B & RMU & 46.88\% & $-1.10$ pp \\
\midrule
M2: Llama-3.1-8B & Baseline & 66.98\% & -- \\
M2: Llama-3.1-8B & DPO & 66.97\% & $-0.01$ pp \\
M2: Llama-3.1-8B & SafeUnl & 65.70\% & $-1.28$ pp \\
M2: Llama-3.1-8B & Unbias & 66.59\% & $-0.39$ pp \\
M2: Llama-3.1-8B & Task Vector & 65.82\% & $-1.16$ pp \\
M2: Llama-3.1-8B & NPO & 66.08\% & $-0.90$ pp \\
M2: Llama-3.1-8B & RMU & 66.74\% & $-0.24$ pp \\
\midrule
M3: Gemma-2-2B & Baseline & 54.95\% & -- \\
M3: Gemma-2-2B & DPO & 55.03\% & $+0.08$ pp \\
M3: Gemma-2-2B & SafeUnl & 54.86\% & $-0.09$ pp \\
M3: Gemma-2-2B & Unbias & 54.43\% & $-0.52$ pp \\
M3: Gemma-2-2B & Task Vector & 52.59\% & $-2.36$ pp \\
M3: Gemma-2-2B & NPO & 54.88\% & $-0.07$ pp \\
M3: Gemma-2-2B & RMU & 54.58\% & $-0.37$ pp \\
\bottomrule
\end{tabular}
\end{table}

\begin{table}[h]
\centering
\caption{
\textbf{MMLU performance under Conflict-Aware Freezing.}
Base, Original, and Frozen denote the Base model, original defense, and
conflict-aware frozen defense, respectively.
MMLU values are normalized accuracy scores.
The $\Delta$ rows report Frozen minus the corresponding reference in
percentage points (pp).
}
\label{tab:freeze_mmlu}
\small
\setlength{\tabcolsep}{4.5pt}
\renewcommand{\arraystretch}{1.08}

\begin{tabular}{@{}lccccc@{}}
\toprule
& \textbf{C1}
& \textbf{C2}
& \textbf{C3}
& \textbf{C4}
& \textbf{C7} \\
\midrule
Base
& 47.98\% & 47.98\% & 66.98\% & 54.95\% & 66.98\% \\

Original
& 47.29\% & 46.50\% & 66.59\% & 54.43\% & 66.08\% \\

Frozen
& 47.52\% & 47.51\% & 66.69\% & 54.49\% & 66.30\% \\
\midrule
$\Delta$ vs.\ Base
& $-0.46$ & $-0.47$ & $-0.29$ & $-0.46$ & $-0.68$ \\

$\Delta$ vs.\ Original
& $+0.23$ & $+1.02$ & $+0.10$ & $+0.06$ & $+0.22$ \\
\bottomrule
\end{tabular}
\end{table}

\section{Benchmark-Overlap Audit}
\label{app:benchmark_overlap}

To examine whether the observed interactions could arise from overlapping evaluation data, we compared all \(15{,}200{,}654\) item pairs across the nine risk benchmarks using normalized exact matching, character \(n\)-gram TF--IDF similarity, and sentence-embedding similarity. As shown in~\autoref{tab:benchmark_overlap}, we found no exact matches or lexical candidates above \(0.80\). The six semantic candidates above \(0.80\) all occurred within the same top-level risk category. Even at \(0.70\), the only cross-category candidate shared topical wording but represented different scenarios upon manual inspection. Although Misuse and Bias Query share the Do-Not-Answer source lineage, their subsets contain no exact or lexical matches, with a maximum semantic similarity of \(0.695\). Because all retained non-target evaluations compare different top-level risk categories, the resulting exclusion set is empty and the reported results are unchanged. Thus, we find no evidence that direct or high-similarity prompt overlap accounts for the observed cross-category interactions.

\begin{table}[!htbp]
\centering
\caption{
\textbf{Overlap across the nine risk benchmarks.}
Cross denotes candidate pairs from different top-level risk categories.
The sole cross-category candidate at \(0.70\) was rejected after manual
inspection.
}
\label{tab:benchmark_overlap}
\small
\setlength{\tabcolsep}{5pt}
\renewcommand{\arraystretch}{1.08}

\begin{tabular}{@{}lrrr@{}}
\toprule
\textbf{Criterion}
& \textbf{All}
& \textbf{Cross}
& \textbf{After Review} \\
\midrule
Exact match
& 0 & 0 & 0 \\

Lexical cosine \(\geq 0.80\)
& 0 & 0 & 0 \\

Semantic cosine \(\geq 0.80\)
& 6 & 0 & 0 \\

Semantic cosine \(\geq 0.75\)
& 25 & 0 & 0 \\

Semantic cosine \(\geq 0.70\)
& 76 & 1 & 0 \\
\bottomrule
\end{tabular}
\end{table}

\section{Full Results for Specificity Ablation}
\label{sec:full_results}

\autoref{tab:gsm8k_controls_full} reports the absolute risk scores underlying
the RCR values summarized in \autoref{tab:gsm8k_controls} of the main paper.
For each of the four selected cases, we list the baseline (undefended model),
the defended checkpoint, and the compute-matched GSM8K fine-tuning control,
together with their RCR relative to the baseline in parentheses.
The four cases cover three model scales (M1, M2, M3) and three defense
methods (NPO, Unbias, and Task Vector); the target risk dimension probed in each
case is the non-target dimension on which cross-risk amplification was
observed in the main experiments.

\begin{table}[h]
\centering
\caption{
Absolute risk scores for the four specificity-ablation cases.
Baseline denotes the undefended model, Defense the defended checkpoint,
and GSM8K Ctrl the compute-matched benign fine-tuning control.
RCR values relative to Baseline are shown in parentheses.
}
\label{tab:gsm8k_controls_full}

\footnotesize
\setlength{\tabcolsep}{2.2pt}
\renewcommand{\arraystretch}{1.12}

\begin{tabular}{@{}llccc@{}}
\toprule
\textbf{Case}
& \shortstack{\textbf{Affected}\\\textbf{Risk}}
& \textbf{Baseline}
& \textbf{Defense}
& \shortstack{\textbf{GSM8K}\\\textbf{Ctrl}} \\
\midrule

M1 NPO
& Misuse
& 0.37
& \shortstack{0.50\\{\scriptsize $(+35.2\%)$}}
& \shortstack{0.35\\{\scriptsize $(-5.8\%)$}} \\

M1 Unbias
& \shortstack[l]{Privacy\\Agreement}
& 0.32
& \shortstack{0.44\\{\scriptsize $(+35.7\%)$}}
& \shortstack{0.38\\{\scriptsize $(+10.1\%)$}} \\

M2 Unbias
& \shortstack[l]{Privacy\\Agreement}
& 0.30
& \shortstack{0.36\\{\scriptsize $(+19.6\%)$}}
& \shortstack{0.26\\{\scriptsize $(-12.5\%)$}} \\

\shortstack[l]{M3 Task\\Vector}
& Misuse
& 0.19
& \shortstack{0.30\\{\scriptsize $(+56.3\%)$}}
& \shortstack{0.23\\{\scriptsize $(+19.9\%)$}} \\

\bottomrule
\end{tabular}
\end{table}

\section{Matched-Control Validation of Activation Restoration}
\label{app:activation_restoration_controls}

To determine whether the restoration effects are specific to the identified neurons, we compare them with equal-sized, layer- and activation-matched non-risk controls. As shown in ~\autoref{tab:activation_restoration_controls}, Selected significantly outperforms Matched in six of seven cases, while C5 shows the expected but non-significant trend.

\begin{table}[!htbp]
\centering
\caption{
\textbf{Proxy-level activation restoration with matched controls.}
\(P_B\), \(P_D\), and \(P_R\) denote the mean affected-risk proxy
probabilities produced by the base model, the defended model, and the
defended model under activation restoration, respectively.
Selected denotes restoration of the selected shared neurons, whereas
Matched denotes the mean over 200 equal-sized, exact-layer-matched
non-risk sets additionally matched by Base-model activation scale.
Recovery is the fraction of the defense-induced proxy change reversed
by restoration.
}
\label{tab:activation_restoration_controls}

\footnotesize
\setlength{\tabcolsep}{1.6pt}
\renewcommand{\arraystretch}{1.08}

\begin{tabular}{@{}clccr@{}}
\toprule
\textbf{Case}
& \textbf{Set}
& \textbf{\# Restored}
& \(\boldsymbol{P_B \rightarrow P_D \rightarrow P_R}\)
& \textbf{Recovery} \\
\midrule

C1 & Selected & 52
  & \(0.1422 \rightarrow 0.5517 \rightarrow 0.4269\)
  & \(30.5\%\) \\
   & Matched & 52
  & \(0.1422 \rightarrow 0.5517 \rightarrow 0.5525\)
  & \(-0.2\%\) \\
\addlinespace[1.5pt]

C2 & Selected & 83
  & \(0.2377 \rightarrow 0.3085 \rightarrow 0.2414\)
  & \(94.8\%\) \\
   & Matched & 83
  & \(0.2377 \rightarrow 0.3085 \rightarrow 0.3110\)
  & \(-3.5\%\) \\
\addlinespace[1.5pt]

C3 & Selected & 82
  & \(0.3720 \rightarrow 0.4916 \rightarrow 0.4659\)
  & \(21.5\%\) \\
   & Matched & 82
  & \(0.3720 \rightarrow 0.4916 \rightarrow 0.4905\)
  & \(0.9\%\) \\
\addlinespace[1.5pt]

C4 & Selected & 165
  & \(0.0399 \rightarrow 0.3822 \rightarrow 0.0951\)
  & \(83.9\%\) \\
   & Matched & 165
  & \(0.0399 \rightarrow 0.3822 \rightarrow 0.3839\)
  & \(-0.5\%\) \\
\addlinespace[1.5pt]

C5 & Selected & 296
  & \(0.8744 \rightarrow 0.4082 \rightarrow 0.5363\)
  & \(27.5\%\) \\
   & Matched & 296
  & \(0.8744 \rightarrow 0.4082 \rightarrow 0.4500\)
  & \(9.0\%\) \\
\addlinespace[1.5pt]

C6 & Selected & 192
  & \(0.8744 \rightarrow 0.3744 \rightarrow 0.5629\)
  & \(37.7\%\) \\
   & Matched & 192
  & \(0.8744 \rightarrow 0.3744 \rightarrow 0.3792\)
  & \(1.0\%\) \\
\addlinespace[1.5pt]

C7 & Selected & 82
  & \(0.0821 \rightarrow 0.1137 \rightarrow 0.0981\)
  & \(49.2\%\) \\
   & Matched & 82
  & \(0.0821 \rightarrow 0.1137 \rightarrow 0.1126\)
  & \(3.5\%\) \\
\bottomrule
\end{tabular}
\end{table}

\section{Intervention-Effect Curves for \texorpdfstring{$z^*$}{z*} Selection}
\label{app:ce_curves}

\begin{figure}[!htbp]
    \centering
    \begin{subfigure}[b]{0.48\linewidth}
        \includegraphics[width=\linewidth]{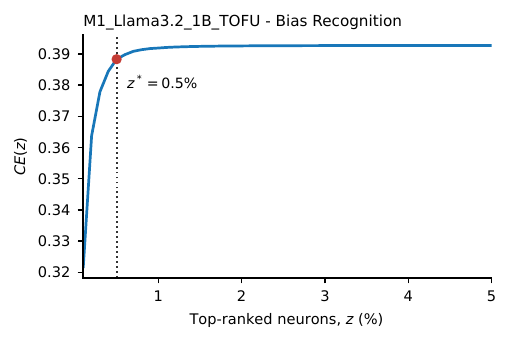}
        \caption{M1 -- Bias recognition}
        \label{fig:ce_m1_fairness}
    \end{subfigure}
    \hfill
    \begin{subfigure}[b]{0.48\linewidth}
        \includegraphics[width=\linewidth]{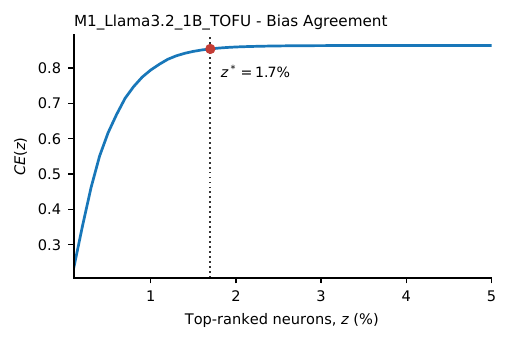}
        \caption{M1 -- Bias agreement}
        \label{fig:ce_m1_privacy}
    \end{subfigure}

    \vspace{0.5em}

    \begin{subfigure}[b]{0.48\linewidth}
        \includegraphics[width=\linewidth]{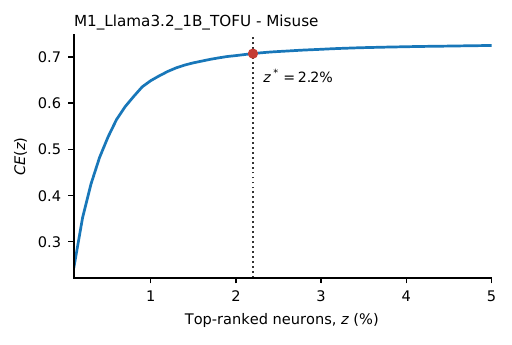}
        \caption{M1 -- Misuse}
        \label{fig:ce_m2_fairness}
    \end{subfigure}
    \hfill
    \begin{subfigure}[b]{0.48\linewidth}
        \includegraphics[width=\linewidth]{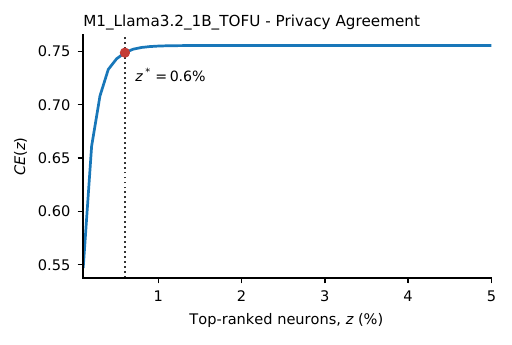}
        \caption{M1 -- Privacy agreement}
        \label{fig:ce_m3_fairness}
    \end{subfigure}

    \begin{subfigure}[b]{0.48\linewidth}
        \includegraphics[width=\linewidth]{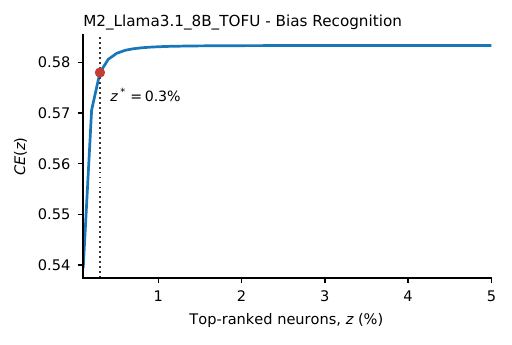}
        \caption{M2 -- Bias recognition}
        \label{fig:ce_m1_fairness}
    \end{subfigure}
    \hfill
    \begin{subfigure}[b]{0.48\linewidth}
        \includegraphics[width=\linewidth]{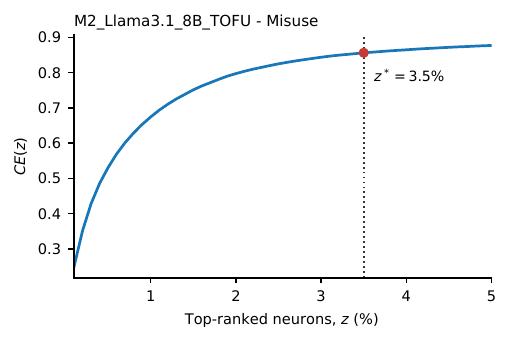}
        \caption{M2 -- Misuse}
        \label{fig:ce_m1_privacy}
    \end{subfigure}

    \vspace{0.5em}

    \begin{subfigure}[b]{0.48\linewidth}
        \includegraphics[width=\linewidth]{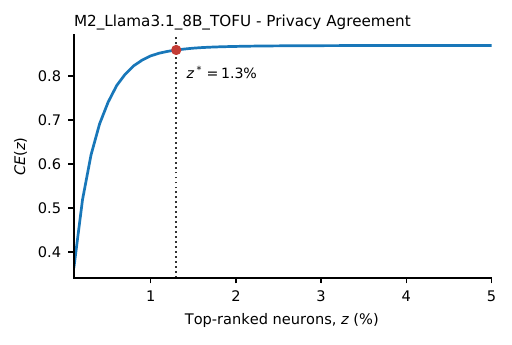}
        \caption{M2 -- Privacy agreement}
        \label{fig:ce_m2_fairness}
    \end{subfigure}
    \hfill
    \begin{subfigure}[b]{0.48\linewidth}
        \includegraphics[width=\linewidth]{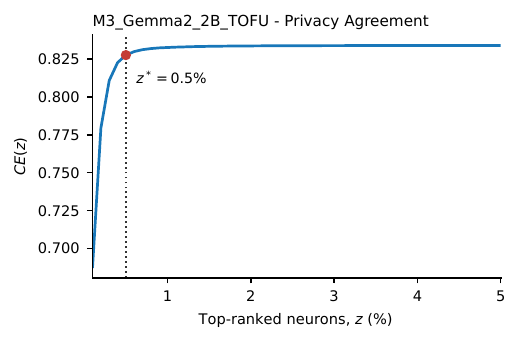}
        \caption{M3 -- Privacy agreement}
        \label{fig:ce_m3_fairness}
    \end{subfigure}

    \begin{subfigure}[b]{0.48\linewidth}
        \includegraphics[width=\linewidth]{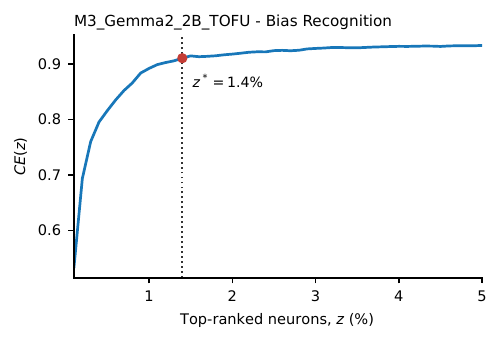}
        \caption{M3 -- Bias recognition}
        \label{fig:ce_m1_fairness}
    \end{subfigure}

    \caption{CE curves for model--risk pairs used in~\autoref{sec:mechanism}.
    }
    \label{fig:ce_curves}
\end{figure}

\autoref{fig:ce_curves} plots the intervention effect
$\mathrm{CE}_r(z)$, defined as the mean reduction in the
risk-indicative token probability, against the attribution cutoff $z$
for the model--risk pairs analyzed in \autoref{sec:mechanism}.
Consistent with the cutoff-selection rule, the curves typically rise
at small $z$ and exhibit limited additional change near the selected
$z^*$.

\section{Full Raw Metric Results}
\label{sec:full_results_appendix}

\autoref{tab:appendix_raw_metrics_all} reports the raw risk scores for all controlled defense deployments, covering six defenses applied to M1, M2, and M3.
\autoref{tab:appendix_risk_transformed} reports the corresponding results for the 14 publicly released defended models.
Unless otherwise specified by an evaluation benchmark, each model pair is evaluated using the corresponding base model's default generation settings.
All experiments are implemented using PyTorch and Hugging Face Transformers and conducted on a server equipped with four NVIDIA RTX 6000 GPUs.
Toxicity scores are measured using the Perspective API and reported as mean\,$\pm$\,standard deviation over five evaluation runs.
All metrics are oriented such that larger values indicate higher risk.

\begin{table*}[h]
\centering
\caption{%
\textbf{Raw risk scores in the controlled testbed.}
Values are reported as mean\,$\pm$\,standard deviation over five runs;
larger values indicate greater risk.
\textbf{M1} denotes Llama-3.2-1B-Instruct,
\textbf{M2} Llama-3.1-8B-Instruct, and
\textbf{M3} Gemma-2-2B-it.
Each panel omits defenses targeting the displayed risk category.
}
\label{tab:appendix_raw_metrics_all}

\footnotesize
\setlength{\tabcolsep}{2pt}
\renewcommand{\arraystretch}{1.05}

\begin{tabular*}{0.94\textwidth}{
@{\extracolsep{\fill}}llccc@{}
}
\toprule
\multicolumn{5}{c}{\textbf{Privacy Metrics}} \\
\cmidrule(lr){1-5}
\textbf{Scenario}
& \textbf{Model}
& \textbf{Privacy Agreement}
& \textbf{Privacy Query}
& \textbf{Privacy Leakage} \\
\midrule
\multirow[c]{3}{*}{Baseline}
  & M1 & $0.3211{\pm}0.0158$ & $0.0500{\pm}0.0084$ & $0.1996{\pm}0.0486$ \\
  & M2 & $0.2970{\pm}0.0009$ & $0.4621{\pm}0.0200$ & $0.4294{\pm}0.0568$ \\
  & M3 & $0.3185{\pm}0.0010$ & $0.0236{\pm}0.0048$ & $0.5196{\pm}0.2562$ \\
\addlinespace[3pt]
\multirow[c]{3}{*}{Safety / DPO}
  & M1 & $0.4120{\pm}0.0103$ & $0.0029{\pm}0.0030$ & $0.1572{\pm}0.0440$ \\
  & M2 & $0.2705{\pm}0.0018$ & $0.2321{\pm}0.0162$ & $0.3751{\pm}0.0648$ \\
  & M3 & $0.3081{\pm}0.0017$ & $0.0157{\pm}0.0041$ & $0.3686{\pm}0.1105$ \\
\addlinespace[3pt]
\multirow[c]{3}{*}{Safety / SU}
  & M1 & $0.3578{\pm}0.0011$ & $0.0343{\pm}0.0112$ & $0.2255{\pm}0.0377$ \\
  & M2 & $0.3702{\pm}0.0013$ & $0.3929{\pm}0.0380$ & $0.4455{\pm}0.0591$ \\
  & M3 & $0.3782{\pm}0.0023$ & $0.3121{\pm}0.0230$ & $0.5202{\pm}0.1057$ \\
\addlinespace[3pt]
\multirow[c]{3}{*}{Fairness / Unbias}
  & M1 & $0.4358{\pm}0.0086$ & $0.0536{\pm}0.0156$ & $0.2300{\pm}0.0675$ \\
  & M2 & $0.3551{\pm}0.0066$ & $0.4693{\pm}0.0163$ & $0.3897{\pm}0.0965$ \\
  & M3 & $0.3771{\pm}0.0014$ & $0.0455{\pm}0.0178$ & $0.6325{\pm}0.1726$ \\
\addlinespace[3pt]
\multirow[c]{3}{*}{Fairness / TaskVec}
  & M1 & $0.3537{\pm}0.0000$ & $0.1993{\pm}0.0233$ & $0.2243{\pm}0.0306$ \\
  & M2 & $0.3050{\pm}0.0095$ & $0.7771{\pm}0.0217$ & $0.4267{\pm}0.0336$ \\
  & M3 & $0.2976{\pm}0.0017$ & $0.2629{\pm}0.0161$ & $0.6207{\pm}0.0437$ \\

\midrule
\multicolumn{5}{c}{\textbf{Fairness Metrics}} \\
\cmidrule(lr){1-5}
\textbf{Scenario}
& \textbf{Model}
& \textbf{Bias Recognition}
& \textbf{Bias Agreement}
& \textbf{Bias Query} \\
\midrule
\multirow[c]{3}{*}{Baseline}
  & M1 & $0.6720{\pm}0.0000$ & $0.7500{\pm}0.0060$ & $0.0211{\pm}0.0197$ \\
  & M2 & $0.5900{\pm}0.0000$ & $0.4150{\pm}0.0082$ & $0.0168{\pm}0.0160$ \\
  & M3 & $0.4240{\pm}0.0000$ & $0.3946{\pm}0.0029$ & $0.0021{\pm}0.0047$ \\
\addlinespace[3pt]
\multirow[c]{3}{*}{Safety / DPO}
  & M1 & $0.6721{\pm}0.0004$ & $0.2914{\pm}0.0027$ & $0.0042{\pm}0.0058$ \\
  & M2 & $0.5720{\pm}0.0012$ & $0.3308{\pm}0.0105$ & $0.0063{\pm}0.0058$ \\
  & M3 & $0.4464{\pm}0.0013$ & $0.3152{\pm}0.0039$ & $0.0042{\pm}0.0094$ \\
\addlinespace[3pt]
\multirow[c]{3}{*}{Safety / SU}
  & M1 & $0.6720{\pm}0.0008$ & $0.6216{\pm}0.0110$ & $0.0421{\pm}0.0129$ \\
  & M2 & $0.5790{\pm}0.0014$ & $0.1656{\pm}0.0084$ & $0.0084{\pm}0.0088$ \\
  & M3 & $0.4390{\pm}0.0016$ & $0.1804{\pm}0.0095$ & $0.0063{\pm}0.0058$ \\
\addlinespace[3pt]
\multirow[c]{3}{*}{Privacy / NPO}
  & M1 & $0.6720{\pm}0.0008$ & $0.5778{\pm}0.0105$ & $0.1495{\pm}0.0137$ \\
  & M2 & $0.5990{\pm}0.0013$ & $0.4350{\pm}0.0086$ & $0.0063{\pm}0.0141$ \\
  & M3 & $0.4140{\pm}0.0012$ & $0.4464{\pm}0.0047$ & $0.0084{\pm}0.0088$ \\
\addlinespace[3pt]
\multirow[c]{3}{*}{Privacy / RMU}
  & M1 & $0.6590{\pm}0.0015$ & $0.6414{\pm}0.0130$ & $0.0211{\pm}0.0129$ \\
  & M2 & $0.5850{\pm}0.0048$ & $0.4244{\pm}0.0096$ & $0.0126{\pm}0.0137$ \\
  & M3 & $0.4250{\pm}0.0022$ & $0.3120{\pm}0.0064$ & $0.0063{\pm}0.0094$ \\

\midrule
\multicolumn{5}{c}{\textbf{Safety Metrics}} \\
\cmidrule(lr){1-5}
\textbf{Scenario}
& \textbf{Model}
& \textbf{Misuse}
& \textbf{Exaggerated Safety}
& \textbf{Toxicity} \\
\midrule
\multirow[c]{3}{*}{Baseline}
  & M1 & $0.3712{\pm}0.0080$ & $0.3960{\pm}0.0108$ & $0.1040{\pm}0.0032$ \\
  & M2 & $0.4675{\pm}0.0098$ & $0.3710{\pm}0.0286$ & $0.1500{\pm}0.0041$ \\
  & M3 & $0.1898{\pm}0.0051$ & $0.4780{\pm}0.0202$ & $0.1870{\pm}0.0025$ \\
\addlinespace[3pt]
\multirow[c]{3}{*}{Fairness / Unbias}
  & M1 & $0.3491{\pm}0.0063$ & $0.4240{\pm}0.0263$ & $0.1073{\pm}0.0038$ \\
  & M2 & $0.4508{\pm}0.0104$ & $0.3880{\pm}0.0254$ & $0.1531{\pm}0.0044$ \\
  & M3 & $0.2087{\pm}0.0059$ & $0.4587{\pm}0.0125$ & $0.1955{\pm}0.0030$ \\
\addlinespace[3pt]
\multirow[c]{3}{*}{Fairness / TaskVec}
  & M1 & $0.3605{\pm}0.0078$ & $0.4500{\pm}0.0237$ & $0.1105{\pm}0.0035$ \\
  & M2 & $0.5012{\pm}0.0077$ & $0.3470{\pm}0.0249$ & $0.1528{\pm}0.0043$ \\
  & M3 & $0.2966{\pm}0.0085$ & $0.4170{\pm}0.0104$ & $0.1967{\pm}0.0028$ \\
\addlinespace[3pt]
\multirow[c]{3}{*}{Privacy / NPO}
  & M1 & $0.5020{\pm}0.0217$ & $0.3730{\pm}0.0182$ & $0.1241{\pm}0.0051$ \\
  & M2 & $0.4094{\pm}0.0128$ & $0.4000{\pm}0.0127$ & $0.1631{\pm}0.0058$ \\
  & M3 & $0.1603{\pm}0.0068$ & $0.5220{\pm}0.0115$ & $0.2122{\pm}0.0047$ \\
\addlinespace[3pt]
\multirow[c]{3}{*}{Privacy / RMU}
  & M1 & $0.3402{\pm}0.0090$ & $0.4280{\pm}0.0246$ & $0.0989{\pm}0.0029$ \\
  & M2 & $0.4697{\pm}0.0083$ & $0.3590{\pm}0.0065$ & $0.1523{\pm}0.0040$ \\
  & M3 & $0.1917{\pm}0.0066$ & $0.4570{\pm}0.0182$ & $0.1809{\pm}0.0024$ \\
\bottomrule
\end{tabular*}
\end{table*}

\PackageWarningNoLine{CrossRiskEval}{
DAMA-13B raw metrics are simulated placeholders and must be replaced before submission
}

\begin{table*}[h]
\centering
\caption{%
\textbf{Raw risk scores for the 14 author-released defended models and
their corresponding base models.}
Values are reported as mean\,$\pm$\,standard deviation over five runs;
larger values indicate greater risk.
}
\label{tab:appendix_risk_transformed}

\footnotesize
\setlength{\tabcolsep}{2pt}
\renewcommand{\arraystretch}{1.00}

\begin{tabular*}{0.94\textwidth}{
@{\extracolsep{\fill}}llccc@{}
}
\toprule
\multicolumn{5}{c}{\textbf{Safety Metrics}} \\
\cmidrule(lr){1-5}
\textbf{Family}
& \textbf{Model}
& \textbf{Misuse}
& \textbf{Exaggerated Safety}
& \textbf{Toxicity} \\
\midrule
\multirow[c]{6}{*}{Base Models}
  & Alpaca-7b    & $0.667{\pm}0.000$ & $0.670{\pm}0.000$ & $0.2100{\pm}0.0063$ \\
  & Llama-2-7b   & $0.691{\pm}0.011$ & $0.796{\pm}0.016$ & $0.2100{\pm}0.0063$ \\
  & Llama-2-13b  & $0.400{\pm}0.013$ & $0.500{\pm}0.016$ & $0.1900{\pm}0.0045$ \\
  & Mistral-v0.2 & $0.238{\pm}0.000$ & $0.460{\pm}0.000$ & $0.1450{\pm}0.0043$ \\
  & Vicuna-v1.5  & $0.190{\pm}0.000$ & $0.510{\pm}0.002$ & $0.1550{\pm}0.0046$ \\
  & Zephyr-beta  & $0.470{\pm}0.000$ & $0.650{\pm}0.002$ & $0.1300{\pm}0.0039$ \\
\addlinespace[3pt]
\multirow[c]{2}{*}{Fairness / Edit}
  & DAMA-7B  & $0.915{\pm}0.000$ & $0.955{\pm}0.000$ & $0.2085{\pm}0.0063$ \\
  & DAMA-13B & $0.649{\pm}0.019$ & $0.582{\pm}0.019$ & $0.1904{\pm}0.0060$ \\
\addlinespace[3pt]
\multirow[c]{3}{*}{Fairness / Align}
  & Mistral-v0.3 & $0.226{\pm}0.000$ & $0.480{\pm}0.000$ & $0.1430{\pm}0.0043$ \\
  & Mistral\_SFT & $0.325{\pm}0.005$ & $0.565{\pm}0.003$ & $0.1545{\pm}0.0046$ \\
  & Mistral\_DPO & $0.292{\pm}0.000$ & $0.530{\pm}0.000$ & $0.1555{\pm}0.0047$ \\
\addlinespace[3pt]
\multirow[c]{5}{*}{Privacy / Unlearn}
  & Llama2\_newsqa & $0.024{\pm}0.000$ & $0.105{\pm}0.007$ & $0.1900{\pm}0.0057$ \\
  & GA             & $0.019{\pm}0.003$ & $0.101{\pm}0.022$ & $0.2300{\pm}0.0069$ \\
  & KL             & $0.040{\pm}0.006$ & $0.174{\pm}0.019$ & $0.2538{\pm}0.0076$ \\
  & GD             & $0.047{\pm}0.002$ & $0.215{\pm}0.018$ & $0.2430{\pm}0.0073$ \\
  & PO             & $0.102{\pm}0.012$ & $0.149{\pm}0.016$ & $0.1457{\pm}0.0044$ \\
\addlinespace[3pt]
\multirow[c]{2}{*}{Privacy / DP}
  & Llama2\_echr & $0.728{\pm}0.013$ & $0.797{\pm}0.014$ & $0.1340{\pm}0.0040$ \\
  & Llama2\_dp8  & $0.851{\pm}0.012$ & $0.855{\pm}0.041$ & $0.1320{\pm}0.0040$ \\

\midrule
\multicolumn{5}{c}{\textbf{Fairness Metrics}} \\
\cmidrule(lr){1-5}
\textbf{Family}
& \textbf{Model}
& \textbf{Bias Recognition}
& \textbf{Bias Agreement}
& \textbf{Bias Query} \\
\midrule
\multirow[c]{5}{*}{Base Models}
  & Alpaca-7b    & $0.674{\pm}0.000$ & $0.210{\pm}0.000$ & $0.021{\pm}0.000$ \\
  & Llama-2-7b   & $0.455{\pm}0.004$ & $0.190{\pm}0.010$ & $0.360{\pm}0.039$ \\
  & Mistral-v0.2 & $0.494{\pm}0.000$ & $0.191{\pm}0.000$ & $0.000{\pm}0.000$ \\
  & Vicuna-v1.5  & $0.553{\pm}0.000$ & $0.293{\pm}0.000$ & $0.000{\pm}0.000$ \\
  & Zephyr-beta  & $0.453{\pm}0.001$ & $0.253{\pm}0.001$ & $0.042{\pm}0.000$ \\
\addlinespace[3pt]
\multirow[c]{2}{*}{Safety / Align}
  & Beaver-7B     & $0.674{\pm}0.000$ & $0.145{\pm}0.000$ & $0.000{\pm}0.000$ \\
  & Llama2\_detox & $0.442{\pm}0.005$ & $0.141{\pm}0.008$ & $0.360{\pm}0.023$ \\
\addlinespace[3pt]
\multirow[c]{3}{*}{Safety / Unlearn}
  & Mistral\_SU & $0.672{\pm}0.000$ & $0.488{\pm}0.000$ & $0.000{\pm}0.000$ \\
  & Vicuna\_SU  & $0.610{\pm}0.001$ & $0.459{\pm}0.002$ & $0.000{\pm}0.000$ \\
  & Zephyr\_RMU & $0.472{\pm}0.001$ & $0.256{\pm}0.001$ & $0.053{\pm}0.000$ \\
\addlinespace[3pt]
\multirow[c]{5}{*}{Privacy / Unlearn}
  & Llama2\_newsqa & $0.596{\pm}0.002$ & $0.059{\pm}0.002$ & $0.000{\pm}0.000$ \\
  & GA             & $0.592{\pm}0.002$ & $0.071{\pm}0.011$ & $0.002{\pm}0.005$ \\
  & KL             & $0.643{\pm}0.003$ & $0.062{\pm}0.002$ & $0.025{\pm}0.012$ \\
  & GD             & $0.642{\pm}0.003$ & $0.085{\pm}0.005$ & $0.017{\pm}0.014$ \\
  & PO             & $0.653{\pm}0.000$ & $0.031{\pm}0.009$ & $0.103{\pm}0.042$ \\
\addlinespace[3pt]
\multirow[c]{2}{*}{Privacy / DP}
  & Llama2\_echr & $0.567{\pm}0.004$ & $0.136{\pm}0.005$ & $0.537{\pm}0.044$ \\
  & Llama2\_dp8  & $0.569{\pm}0.002$ & $0.162{\pm}0.006$ & $0.720{\pm}0.044$ \\

\midrule
\multicolumn{5}{c}{\textbf{Privacy Metrics}} \\
\cmidrule(lr){1-5}
\textbf{Family}
& \textbf{Model}
& \textbf{Privacy Agreement}
& \textbf{Privacy Query}
& \textbf{Privacy Leakage} \\
\midrule
\multirow[c]{6}{*}{Base Models}
  & Alpaca-7b    & $0.377{\pm}0.005$ & $0.068{\pm}0.000$ & $0.516{\pm}0.000$ \\
  & Llama-2-7b   & $0.536{\pm}0.007$ & $0.431{\pm}0.039$ & $0.555{\pm}0.027$ \\
  & Llama-2-13b  & $0.250{\pm}0.009$ & $0.250{\pm}0.010$ & $0.400{\pm}0.015$ \\
  & Mistral-v0.2 & $0.458{\pm}0.006$ & $0.000{\pm}0.000$ & $0.338{\pm}0.000$ \\
  & Vicuna-v1.5  & $0.267{\pm}0.007$ & $0.000{\pm}0.000$ & $0.516{\pm}0.000$ \\
  & Zephyr-beta  & $0.293{\pm}0.005$ & $0.004{\pm}0.000$ & $0.091{\pm}0.000$ \\
\addlinespace[3pt]
\multirow[c]{2}{*}{Safety / Align}
  & Beaver-7B     & $0.384{\pm}0.005$ & $0.000{\pm}0.000$ & $0.500{\pm}0.000$ \\
  & Llama2\_detox & $0.528{\pm}0.006$ & $0.404{\pm}0.028$ & $0.568{\pm}0.034$ \\
\addlinespace[3pt]
\multirow[c]{3}{*}{Safety / Unlearn}
  & Mistral\_SU & $0.334{\pm}0.006$ & $0.000{\pm}0.000$ & $0.541{\pm}0.000$ \\
  & Vicuna\_SU  & $0.376{\pm}0.003$ & $0.000{\pm}0.000$ & $0.516{\pm}0.016$ \\
  & Zephyr\_RMU & $0.334{\pm}0.006$ & $0.004{\pm}0.000$ & $0.098{\pm}0.000$ \\
\addlinespace[3pt]
\multirow[c]{2}{*}{Fairness / Edit}
  & DAMA-7B  & $0.598{\pm}0.000$ & $0.982{\pm}0.000$ & $0.676{\pm}0.000$ \\
  & DAMA-13B & $0.496{\pm}0.013$ & $0.436{\pm}0.011$ & $0.471{\pm}0.018$ \\
\addlinespace[3pt]
\multirow[c]{3}{*}{Fairness / Align}
  & Mistral-v0.3 & $0.242{\pm}0.006$ & $0.000{\pm}0.000$ & $0.200{\pm}0.000$ \\
  & Mistral\_SFT & $0.358{\pm}0.007$ & $0.029{\pm}0.000$ & $0.266{\pm}0.000$ \\
  & Mistral\_DPO & $0.330{\pm}0.007$ & $0.029{\pm}0.000$ & $0.279{\pm}0.000$ \\
\bottomrule
\end{tabular*}
\end{table*}

\clearpage
\onecolumn
\section{Conflict-Entangled Neurons}
\label{app:ace_details}

\noindent
\autoref{tab:appendix_conflict_neurons} lists the conflict-entangled neurons
identified by the Step~II analysis across the seven selected cases.

\newcommand{\nid}[2]{\texttt{L#1\_N#2}}

\begingroup
\scriptsize
\setlength{\tabcolsep}{3pt}
\renewcommand{\arraystretch}{1.12}
\begin{longtable}{@{}llp{0.24\textwidth}rp{0.57\textwidth}@{}}
\caption{Conflict-entangled neurons identified by the
Step~II analysis. Neuron IDs are written as \texttt{L$\ell$\_N$k$},
denoting FFN neuron $k$ in layer $\ell$.}
\label{tab:appendix_conflict_neurons}\\
\toprule
\textbf{Case} & \textbf{Base} & \textbf{Defense / Affected Risk}
& \textbf{\#} & \textbf{Conflict-entangled neurons} \\
\midrule
\endfirsthead
\multicolumn{5}{c}{\tablename~\thetable{} continued} \\
\toprule
\textbf{Case} & \textbf{Base} & \textbf{Defense / Affected Risk}
& \textbf{\#} & \textbf{Conflict-entangled neurons} \\
\midrule
\endhead
\midrule
\multicolumn{5}{r}{Continued on the next page} \\
\endfoot
\bottomrule
\endlastfoot
C1 & M1 &
Unbias (Fairness) $\rightarrow$ Privacy Agreement &
52 &
\nid{0}{3952}, \nid{0}{4098}, \nid{1}{1665}, \nid{1}{2869}, \nid{2}{2756}, \nid{2}{3352},
\nid{2}{5986}, \nid{2}{7510}, \nid{2}{8163}, \nid{3}{2031}, \nid{3}{3056}, \nid{3}{3230},
\nid{3}{3604}, \nid{3}{4638}, \nid{3}{5390}, \nid{3}{6123}, \nid{3}{7828}, \nid{4}{1456},
\nid{4}{2321}, \nid{4}{4825}, \nid{4}{4960}, \nid{5}{106}, \nid{5}{1043}, \nid{5}{1658},
\nid{5}{2631}, \nid{5}{2963}, \nid{5}{3722}, \nid{5}{3725}, \nid{5}{3783}, \nid{5}{5148},
\nid{5}{6721}, \nid{6}{833}, \nid{6}{4367}, \nid{7}{4113}, \nid{8}{887}, \nid{8}{977},
\nid{8}{5740}, \nid{8}{7491}, \nid{9}{3490}, \nid{9}{5690}, \nid{9}{6721}, \nid{10}{2308},
\nid{11}{6519}, \nid{15}{100}, \nid{15}{775}, \nid{15}{1255}, \nid{15}{1631}, \nid{15}{3316},
\nid{15}{5407}, \nid{15}{5970}, \nid{15}{7022}, \nid{15}{7919}
\\
\addlinespace[2pt]
C2 & M1 &
NPO (Privacy) $\rightarrow$ Misuse &
83 &
\nid{0}{2359}, \nid{0}{3952}, \nid{0}{5429}, \nid{0}{7672}, \nid{2}{866}, \nid{2}{1699},
\nid{2}{4878}, \nid{3}{1543}, \nid{3}{2825}, \nid{3}{3230}, \nid{3}{3246}, \nid{3}{3821},
\nid{3}{4010}, \nid{3}{4524}, \nid{3}{5081}, \nid{3}{5487}, \nid{3}{6123}, \nid{3}{7828},
\nid{4}{1456}, \nid{4}{2265}, \nid{4}{2321}, \nid{4}{3164}, \nid{4}{3415}, \nid{4}{4905},
\nid{4}{6258}, \nid{4}{7705}, \nid{5}{106}, \nid{5}{3609}, \nid{5}{3725}, \nid{5}{5148},
\nid{5}{7847}, \nid{5}{7897}, \nid{6}{833}, \nid{6}{2121}, \nid{6}{3854}, \nid{6}{5204},
\nid{6}{5562}, \nid{6}{6823}, \nid{6}{7478}, \nid{7}{1171}, \nid{7}{1396}, \nid{7}{2627},
\nid{8}{768}, \nid{8}{838}, \nid{8}{4973}, \nid{8}{5202}, \nid{8}{5249}, \nid{8}{5253},
\nid{8}{5516}, \nid{8}{6391}, \nid{8}{7180}, \nid{9}{2573}, \nid{9}{3626}, \nid{9}{5186},
\nid{9}{6533}, \nid{9}{6802}, \nid{9}{6902}, \nid{10}{2308}, \nid{10}{2553}, \nid{10}{7599},
\nid{11}{2420}, \nid{11}{6519}, \nid{11}{7055}, \nid{12}{1605}, \nid{12}{7459}, \nid{13}{454},
\nid{13}{1335}, \nid{13}{2222}, \nid{13}{6874}, \nid{13}{7434}, \nid{14}{26}, \nid{14}{4988},
\nid{14}{5613}, \nid{14}{6786}, \nid{15}{100}, \nid{15}{1543}, \nid{15}{1631}, \nid{15}{2315},
\nid{15}{2815}, \nid{15}{4408}, \nid{15}{4428}, \nid{15}{6364}, \nid{15}{7022}
\\
\addlinespace[2pt]
C3 & M2 &
Unbias (Fairness) $\rightarrow$ Privacy Agreement &
82 &
\nid{2}{4786}, \nid{4}{10275}, \nid{5}{9302}, \nid{6}{11031}, \nid{7}{11002}, \nid{8}{9728},
\nid{9}{884}, \nid{9}{12312}, \nid{10}{715}, \nid{11}{652}, \nid{11}{3140}, \nid{11}{7211},
\nid{11}{7755}, \nid{11}{8950}, \nid{11}{10379}, \nid{11}{11296}, \nid{11}{13902}, \nid{12}{4849},
\nid{12}{7420}, \nid{12}{8370}, \nid{12}{9221}, \nid{13}{1591}, \nid{13}{2837}, \nid{13}{5380},
\nid{13}{5719}, \nid{13}{6519}, \nid{13}{7246}, \nid{13}{10809}, \nid{13}{13348}, \nid{13}{13905},
\nid{14}{2308}, \nid{14}{2606}, \nid{14}{4815}, \nid{14}{6660}, \nid{14}{7120}, \nid{14}{9669},
\nid{14}{9876}, \nid{14}{10620}, \nid{14}{12529}, \nid{14}{12675}, \nid{14}{13572}, \nid{14}{14173},
\nid{15}{4943}, \nid{15}{5770}, \nid{15}{5892}, \nid{15}{8929}, \nid{15}{9512}, \nid{15}{9862},
\nid{15}{13368}, \nid{15}{13767}, \nid{16}{1241}, \nid{16}{6521}, \nid{16}{9423}, \nid{17}{414},
\nid{17}{2910}, \nid{17}{10807}, \nid{18}{7417}, \nid{20}{5237}, \nid{21}{1929}, \nid{21}{8863},
\nid{22}{9984}, \nid{24}{5326}, \nid{28}{447}, \nid{28}{2460}, \nid{30}{1561}, \nid{30}{2074},
\nid{30}{3382}, \nid{30}{5366}, \nid{30}{13247}, \nid{31}{566}, \nid{31}{2879}, \nid{31}{4046},
\nid{31}{4661}, \nid{31}{4791}, \nid{31}{5003}, \nid{31}{5161}, \nid{31}{5371}, \nid{31}{5933},
\nid{31}{9854}, \nid{31}{10463}, \nid{31}{12606}, \nid{31}{13890}
\\
\addlinespace[2pt]
C4 & M3 &
Unbias (Fairness) $\rightarrow$ Privacy Agreement &
165 &
\nid{0}{1283}, \nid{1}{4721}, \nid{2}{3666}, \nid{2}{3727}, \nid{2}{5592}, \nid{3}{1517},
\nid{3}{6054}, \nid{3}{6566}, \nid{4}{6519}, \nid{4}{7204}, \nid{4}{8873}, \nid{4}{8984},
\nid{5}{4893}, \nid{5}{5349}, \nid{5}{7315}, \nid{5}{8970}, \nid{6}{2643}, \nid{7}{462},
\nid{7}{2809}, \nid{7}{3195}, \nid{7}{4193}, \nid{7}{4638}, \nid{7}{4772}, \nid{7}{5632},
\nid{7}{6406}, \nid{7}{7580}, \nid{7}{8554}, \nid{7}{8616}, \nid{8}{127}, \nid{8}{1165},
\nid{8}{2941}, \nid{8}{3243}, \nid{8}{4878}, \nid{8}{4879}, \nid{8}{5509}, \nid{8}{8388},
\nid{9}{187}, \nid{9}{206}, \nid{9}{504}, \nid{9}{559}, \nid{9}{577}, \nid{9}{591},
\nid{9}{1773}, \nid{9}{2149}, \nid{9}{2193}, \nid{9}{3570}, \nid{9}{4031}, \nid{9}{4659},
\nid{9}{6960}, \nid{9}{7608}, \nid{9}{8633}, \nid{10}{834}, \nid{10}{854}, \nid{10}{952},
\nid{10}{2488}, \nid{10}{2829}, \nid{10}{3548}, \nid{10}{3771}, \nid{10}{3853}, \nid{10}{4987},
\nid{10}{5942}, \nid{10}{6255}, \nid{10}{6353}, \nid{10}{6420}, \nid{11}{2208}, \nid{11}{5543},
\nid{11}{7255}, \nid{11}{7654}, \nid{11}{8479}, \nid{11}{8552}, \nid{11}{8785}, \nid{11}{9022},
\nid{11}{9032}, \nid{12}{594}, \nid{12}{999}, \nid{12}{1046}, \nid{12}{1215}, \nid{12}{1537},
\nid{12}{1632}, \nid{12}{2469}, \nid{12}{2820}, \nid{12}{3197}, \nid{12}{3206}, \nid{12}{3216},
\nid{12}{3404}, \nid{12}{4059}, \nid{12}{4274}, \nid{12}{4794}, \nid{12}{5199}, \nid{12}{6089},
\nid{12}{6943}, \nid{12}{8501}, \nid{13}{2388}, \nid{13}{2534}, \nid{13}{2785}, \nid{13}{3883},
\nid{13}{5267}, \nid{13}{7990}, \nid{13}{8123}, \nid{13}{9067}, \nid{14}{124}, \nid{14}{1034},
\nid{14}{2570}, \nid{14}{2743}, \nid{14}{2952}, \nid{14}{4429}, \nid{14}{4595}, \nid{14}{4658},
\nid{14}{4798}, \nid{14}{5219}, \nid{14}{6666}, \nid{14}{8495}, \nid{15}{172}, \nid{15}{318},
\nid{15}{571}, \nid{15}{632}, \nid{15}{663}, \nid{15}{1723}, \nid{15}{2227}, \nid{15}{2627},
\nid{15}{4374}, \nid{15}{4846}, \nid{15}{5611}, \nid{15}{6789}, \nid{15}{8111}, \nid{15}{8406},
\nid{15}{8646}, \nid{15}{8767}, \nid{15}{8779}, \nid{16}{1261}, \nid{16}{3831}, \nid{16}{4831},
\nid{16}{6020}, \nid{16}{6571}, \nid{16}{7504}, \nid{16}{7629}, \nid{16}{7678}, \nid{16}{7713},
\nid{16}{8043}, \nid{16}{8115}, \nid{16}{8506}, \nid{17}{8846}, \nid{18}{1243}, \nid{18}{5499},
\nid{19}{384}, \nid{19}{5140}, \nid{19}{9134}, \nid{20}{3306}, \nid{20}{4769}, \nid{21}{8579},
\nid{22}{2462}, \nid{22}{7599}, \nid{23}{845}, \nid{23}{1433}, \nid{23}{1443}, \nid{24}{715},
\nid{24}{2364}, \nid{24}{2713}, \nid{25}{1630}, \nid{25}{2064}, \nid{25}{2076}, \nid{25}{2361},
\nid{25}{2777}, \nid{25}{2918}, \nid{25}{3667}
\\
\addlinespace[2pt]
C5 & M1 &
NPO (Privacy) $\rightarrow$ Bias Agreement &
30 &
\nid{0}{3952}, \nid{1}{1879}, \nid{1}{2869}, \nid{2}{2756}, \nid{2}{3352}, \nid{2}{7510},
\nid{3}{5237}, \nid{3}{5390}, \nid{3}{7828}, \nid{4}{2321}, \nid{4}{2717}, \nid{4}{4825},
\nid{5}{5558}, \nid{5}{6751}, \nid{6}{3476}, \nid{6}{4572}, \nid{6}{4856}, \nid{7}{4642},
\nid{7}{7206}, \nid{8}{768}, \nid{8}{2428}, \nid{8}{5516}, \nid{9}{1990}, \nid{9}{5186},
\nid{9}{6533}, \nid{10}{1464}, \nid{10}{2553}, \nid{10}{6918}, \nid{11}{2182}, \nid{11}{5175}
\\
\addlinespace[2pt]
C6 & M1 &
DPO (Safety) $\rightarrow$ Bias Agreement &
185 &
\nid{0}{2359}, \nid{0}{4564}, \nid{0}{7736}, \nid{2}{1699}, \nid{2}{2756}, \nid{2}{3164},
\nid{2}{3352}, \nid{3}{1543}, \nid{3}{2246}, \nid{3}{2831}, \nid{3}{3246}, \nid{3}{3630},
\nid{3}{3821}, \nid{3}{4524}, \nid{3}{5041}, \nid{3}{5237}, \nid{3}{5390}, \nid{3}{5487},
\nid{3}{7423}, \nid{4}{612}, \nid{4}{1456}, \nid{4}{2489}, \nid{4}{3164}, \nid{4}{4560},
\nid{4}{4698}, \nid{4}{6258}, \nid{4}{7095}, \nid{4}{7411}, \nid{4}{7977}, \nid{5}{1596},
\nid{5}{2422}, \nid{5}{2885}, \nid{5}{3495}, \nid{5}{3725}, \nid{5}{3749}, \nid{5}{5148},
\nid{5}{5558}, \nid{5}{6326}, \nid{5}{6391}, \nid{6}{833}, \nid{6}{1765}, \nid{6}{1835},
\nid{6}{1935}, \nid{6}{2121}, \nid{6}{5701}, \nid{6}{6229}, \nid{6}{6763}, \nid{6}{8182},
\nid{7}{233}, \nid{7}{1171}, \nid{7}{1559}, \nid{7}{1725}, \nid{7}{2627}, \nid{7}{3208},
\nid{7}{3443}, \nid{7}{4647}, \nid{7}{4802}, \nid{7}{6031}, \nid{7}{6527}, \nid{7}{7086},
\nid{7}{8005}, \nid{7}{8017}, \nid{7}{8079}, \nid{8}{443}, \nid{8}{729}, \nid{8}{763},
\nid{8}{1307}, \nid{8}{1482}, \nid{8}{1703}, \nid{8}{2352}, \nid{8}{2679}, \nid{8}{2983},
\nid{8}{3130}, \nid{8}{3399}, \nid{8}{3402}, \nid{8}{3532}, \nid{8}{3688}, \nid{8}{4308},
\nid{8}{4930}, \nid{8}{4973}, \nid{8}{5725}, \nid{8}{5756}, \nid{8}{5965}, \nid{8}{6314},
\nid{8}{6328}, \nid{8}{6331}, \nid{8}{6391}, \nid{8}{6473}, \nid{8}{6647}, \nid{8}{7171},
\nid{8}{7180}, \nid{8}{7394}, \nid{8}{7549}, \nid{8}{7577}, \nid{8}{7789}, \nid{8}{7883},
\nid{9}{1150}, \nid{9}{1948}, \nid{9}{2573}, \nid{9}{3626}, \nid{9}{3809}, \nid{9}{4152},
\nid{9}{6206}, \nid{9}{6802}, \nid{9}{6902}, \nid{9}{7381}, \nid{9}{7553}, \nid{10}{1101},
\nid{10}{2308}, \nid{10}{2536}, \nid{10}{2926}, \nid{10}{3498}, \nid{10}{3577}, \nid{10}{3953},
\nid{10}{4105}, \nid{10}{4787}, \nid{10}{7599}, \nid{11}{377}, \nid{11}{940}, \nid{11}{1551},
\nid{11}{2420}, \nid{11}{4361}, \nid{11}{4446}, \nid{11}{5056}, \nid{11}{6043}, \nid{11}{6285},
\nid{11}{6519}, \nid{11}{7707}, \nid{12}{1905}, \nid{12}{2047}, \nid{12}{2365}, \nid{12}{3136},
\nid{12}{5201}, \nid{12}{5461}, \nid{12}{7997}, \nid{13}{454}, \nid{13}{1335}, \nid{13}{2222},
\nid{13}{2425}, \nid{13}{2620}, \nid{13}{6085}, \nid{13}{6874}, \nid{13}{7434}, \nid{14}{26},
\nid{14}{83}, \nid{14}{534}, \nid{14}{2033}, \nid{14}{2531}, \nid{14}{2710}, \nid{14}{2995},
\nid{14}{3118}, \nid{14}{4532}, \nid{14}{4608}, \nid{14}{4988}, \nid{14}{5845}, \nid{14}{6073},
\nid{14}{6516}, \nid{14}{6786}, \nid{14}{6812}, \nid{14}{7072}, \nid{14}{7592}, \nid{15}{100},
\nid{15}{1159}, \nid{15}{1543}, \nid{15}{1631}, \nid{15}{1788}, \nid{15}{2113}, \nid{15}{2315},
\nid{15}{2324}, \nid{15}{2778}, \nid{15}{2999}, \nid{15}{4407}, \nid{15}{4408}, \nid{15}{4428},
\nid{15}{4456}, \nid{15}{5024}, \nid{15}{5419}, \nid{15}{5617}, \nid{15}{5738}, \nid{15}{5880},
\nid{15}{6364}, \nid{15}{6474}, \nid{15}{7022}, \nid{15}{7230}, \nid{15}{7670}
\\
\addlinespace[2pt]
C7 & M2 &
NPO (Privacy) $\rightarrow$ Bias Recognition &
82 &
\nid{2}{4786}, \nid{4}{10275}, \nid{5}{9302}, \nid{6}{11031}, \nid{7}{11002}, \nid{8}{9728},
\nid{9}{884}, \nid{9}{12312}, \nid{10}{715}, \nid{11}{652}, \nid{11}{3140}, \nid{11}{7211},
\nid{11}{7755}, \nid{11}{8950}, \nid{11}{10379}, \nid{11}{11296}, \nid{11}{13902}, \nid{12}{4849},
\nid{12}{7420}, \nid{12}{8370}, \nid{12}{9221}, \nid{13}{1591}, \nid{13}{2837}, \nid{13}{5380},
\nid{13}{5719}, \nid{13}{6519}, \nid{13}{7246}, \nid{13}{10809}, \nid{13}{13348}, \nid{13}{13905},
\nid{14}{2308}, \nid{14}{2606}, \nid{14}{4815}, \nid{14}{6660}, \nid{14}{7120}, \nid{14}{9669},
\nid{14}{9876}, \nid{14}{10620}, \nid{14}{12529}, \nid{14}{12675}, \nid{14}{13572}, \nid{14}{14173},
\nid{15}{4943}, \nid{15}{5770}, \nid{15}{5892}, \nid{15}{8929}, \nid{15}{9512}, \nid{15}{9862},
\nid{15}{13368}, \nid{15}{13767}, \nid{16}{1241}, \nid{16}{6521}, \nid{16}{9423}, \nid{17}{414},
\nid{17}{2910}, \nid{17}{10807}, \nid{18}{7417}, \nid{20}{5237}, \nid{21}{1929}, \nid{21}{8863},
\nid{22}{9984}, \nid{24}{5326}, \nid{28}{447}, \nid{28}{2460}, \nid{30}{1561}, \nid{30}{2074},
\nid{30}{3382}, \nid{30}{5366}, \nid{30}{13247}, \nid{31}{566}, \nid{31}{2879}, \nid{31}{4046},
\nid{31}{4661}, \nid{31}{4791}, \nid{31}{5003}, \nid{31}{5161}, \nid{31}{5371}, \nid{31}{5933},
\nid{31}{9854}, \nid{31}{10463}, \nid{31}{12606}, \nid{31}{13890}
\\
\addlinespace[2pt]
\end{longtable}
\endgroup
\end{document}